\documentclass[conference]{IEEEtran}
\IEEEoverridecommandlockouts

\usepackage{cite}
\usepackage{amsmath,amssymb,amsfonts}
\usepackage{algorithmic}
\usepackage{graphicx}
\usepackage{textcomp}
\usepackage{xcolor}
\usepackage[hyphens]{url}
\usepackage{fancyhdr}
\usepackage[bookmarks=true,breaklinks=true,letterpaper=true,colorlinks,citecolor=blue,linkcolor=blue,urlcolor=blue]{hyperref}
\def\BibTeX{{\rm B\kern-.05em{\sc i\kern-.025em b}\kern-.08em
    T\kern-.1667em\lower.7ex\hbox{E}\kern-.125emX}}

\usepackage{physics}
\usepackage{algorithmic}
\usepackage{multirow}
\usepackage{amsthm}
\usepackage{enumitem}
\usepackage{tikz}
\usepackage[ruled]{algorithm2e}

\newtheorem{thm}{Theorem}

\newcommand*\bcircled[1]{\tikz[baseline=(char.base)]{
            \node[shape=circle,draw,inner sep=0.5pt, fill=black, text=white] (char) {#1};}}

\newcommand{\ignore}[1]{}

\definecolor{oldaliceblue}{HTML}{070A52}
\definecolor{aliceblue}{rgb}{0.14, 0.01, 0.8}
\definecolor{OliveGreen}{HTML}{1A4D2E}
\definecolor{orange}{HTML}{1A4D2E}
\usepackage{xcolor}
\usepackage{tcolorbox}
\tcbset{width=1\columnwidth,boxrule=1pt,colback=aliceblue,arc=1pt,left=2pt,right=2pt,boxsep=0.5pt}

\newtcolorbox{hintbox}[2][]
{
  colframe = oldaliceblue!100,
  colback  = oldaliceblue!5,
  boxsep=2pt,
  width=\dimexpr\columnwidth\relax, 
  coltitle = oldaliceblue!20!black,
  title    = #2,
  #1,
}
    
\begin{document}

\title{\huge{Flag-Proxy Networks: Overcoming the Architectural, Scheduling and Decoding Obstacles of Quantum LDPC Codes}
}

\makeatletter
\newcommand{\linebreakand}{%
  \end{@IEEEauthorhalign}
  \hfill\mbox{}\par
  \mbox{}\hfill\begin{@IEEEauthorhalign}
}
\makeatother

\author{\IEEEauthorblockN{Suhas Vittal}
\IEEEauthorblockA{\textit{Georgia Institute of Technology} \\
Atlanta, GA, United States\\
suhaskvittal@gatech.edu}
\and
\IEEEauthorblockN{Ali Javadi-Abhari}
\IEEEauthorblockA{\textit{IBM T.J. Watson Research Center} \\
Yorktown Heights, NY, United States \\
ali.javadi@ibm.com}
\and
\IEEEauthorblockN{Andrew W. Cross}
\IEEEauthorblockA{\textit{IBM T.J. Watson Research Center} \\
Yorktown Heights, NY, United States \\
awcross@us.ibm.com}
\linebreakand
\IEEEauthorblockN{Lev S. Bishop}
\IEEEauthorblockA{\textit{IBM T.J. Watson Research Center} \\
Yorktown Heights, NY, United States \\
lsbishop@us.ibm.com}
\and
\IEEEauthorblockN{Moinuddin Qureshi}
\IEEEauthorblockA{\textit{Georgia Institute of Technology} \\
Atlanta, GA, United States\\
moin@gatech.edu}
}

\maketitle

\begin{abstract}

Quantum error correction is necessary for achieving exponential speedups on important applications. The planar surface code has remained the most studied error-correcting code for the last two decades because of its relative simplicity. However, encoding a singular logical qubit with the planar surface code requires physical qubits quadratic in the code distance~($d$), making it space-inefficient for the large-distance codes necessary for promising applications. Thus, {\em Quantum Low-Density Parity-Check (QLDPC)} have emerged as an alternative to the planar surface code but require a higher degree of connectivity. Furthermore, the problems of fault-tolerant syndrome extraction and decoding are understudied for these codes and also remain obstacles to their usage.

In this paper, we consider two under-studied families of QLDPC codes: hyperbolic surface codes and hyperbolic color codes. We tackle the three challenges mentioned above as follows. {\em First}, we propose {\em Flag-Proxy Networks (FPNs)}, a generalizable architecture for quantum codes that achieves low connectivity through flag and proxy qubits. {\em Second}, we propose a {\em greedy syndrome extraction scheduling} algorithm for general quantum codes and further use this algorithm for fault-tolerant syndrome extraction on FPNs. {\em Third}, we present two decoders that leverage flag measurements to decode the hyperbolic codes accurately. Our work finds that degree-4 FPNs of the hyperbolic surface and color codes are respectively $2.9\times$ and $5.5\times$ more space-efficient than the $d = 5$ planar surface code, and become even more space-efficient when considering higher distances. The hyperbolic codes also have error rates comparable to their planar counterparts.

\end{abstract}

\section{Introduction}

Quantum error correction is the most promising path towards realizing exponential speedups for applications in quantum chemistry and cryptanalysis~\cite{childs2018firstqsim, campbell2019qalgsconstraintsat, kivlichan2020qsimelectrons, shor1999polynomial, harrow2009hhl, reiher2017nitrogenfixation}. Error-corrected quantum computers leverage quantum error correction by encoding multiple noisy \textit{physical} qubits into a \textit{logical block} containing fewer error-resilient logical qubits. The logical qubits encoded by quantum error correcting codes have better fidelity than their constituent physical qubits, provided the physical error rate is low enough (e.g., 0.1\%).

\begin{figure*}
    \centering
    \includegraphics[width=\textwidth]{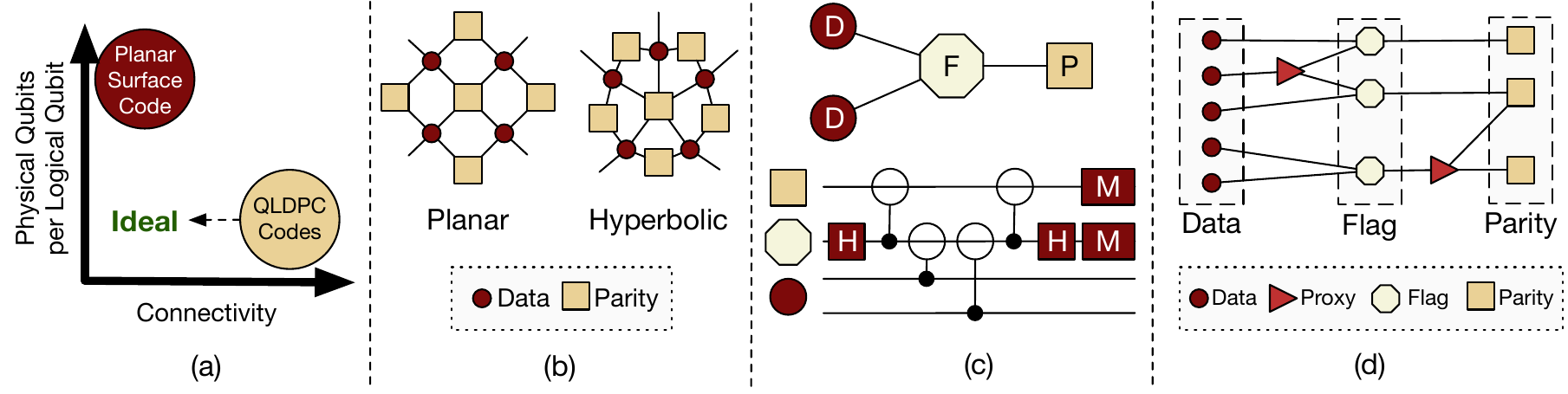}
    \caption{(a)~Tradeoff between efficiency and connectivity amongst quantum error correcting codes. (b)~Local structures for the planar and hyperbolic surface codes. (c)~Syndrome extraction with a flag qubit is measured to detect errors on the data qubits. (d)~The proposed Flag-Proxy Network architecture.}
    \label{fig:intro}
\end{figure*}

In this paper, we focus on superconducting quantum computers. Quantum error-correcting codes implemented on superconducting quantum computers arrange physical qubits into \textit{data qubits}, which maintain the logical state, and \textit{parity qubits}, which detect $X$ and $Z$ errors. Leveraging these parity qubits requires executing a quantum \textit{syndrome extraction circuit}, which entangles the parity qubits with neighboring data qubits. Subsequently, these parity qubits are measured, yielding a bitstring known as a \textit{syndrome}. The syndrome is sent to a \textit{decoder}, which identifies errors on the data qubits. However, since syndromes are unreliable as syndrome extraction itself is erroneous, the decoder must analyze $d$ consecutive rounds of syndromes to identify errors accurately. Ideally, a distance $d$ quantum error-correcting code should correct up to $(d-1)/2$ errors within $d$ syndrome extraction rounds; increasing $d$ exponentially suppresses error. However, circuit errors can harm the \textit{effective distance} of the code, preventing $(d-1)/2$ errors from being corrected. A syndrome extraction circuit is \textit{fault-tolerant} if the effective distance is sufficiently high such that errors remain exponentially suppressed with increasing $d$.

The \textit{planar surface code} has remained the focus of error correction research for the last two decades. We attribute this dominance to three reasons:
\begin{enumerate}
[leftmargin=0.5cm,itemindent=0.4cm,labelwidth=\itemindent,labelsep=0.05cm, align=left, itemsep=0.1cm, listparindent=0.5cm, topsep=0.1cm]
    \item The planar surface code requires grid connectivity, which is easy to fabricate using superconducting qubits. For this reason, the planar surface code has been realized on multiple quantum processors~\cite{krinner2022distance3demonstration, google2023suppressing, google2024distance7sc}. In contrast, more densely connected processors are hard to fabricate due to frequency crowding and crosstalk\cite{li2021qswhwcodesign, murali2020crosstalk}.
    \item Fault-tolerant syndrome extraction for the planar surface code is enabled by diligent CNOT scheduling~\cite{tomita2014surfacecodewithrealisticnoise}. In contrast, fault-tolerant syndrome extraction with other quantum codes requires additional physical overheads beyond data and parity qubits~\cite{chamberland2020heavycodes, chamberland2020colorcodeswithflags, chao2018flagqubits}.
    \item The planar surface code can be decoded with the \textit{Minimum Weight Perfect Matching (MWPM)} algorithm, whose implementations are fast and readily available~\cite{higgott2022pymatching, higgott2023sparseblossom}.
\end{enumerate}
However, the planar surface code is space-inefficient as any amount of redundancy encodes a singular logical qubit. For any promising applications, implementing a single logical qubit at a near-term error rate of $p = 10^{-3}$ consumes at least $1000$ physical qubits~\cite{litinski2019gameofsurfacecodes}. Since many promising applications require 1000s of high-fidelity program qubits~\cite{gidney2021rsafactorization}, error-corrected quantum computers exclusively using the planar surface code will require millions of physical qubits to provide quantum advantage. These overheads are undesirable given the engineering challenges of building very large quantum computers. Ideally, we desire alternative quantum codes with better efficiency and comparable error rates.

\vspace{0.05in}
\noindent
\textbf{Moving Away from the Planar Surface Code:} 
\textit{Quantum Low-Density Parity Check (QLDPC)} codes have emerged as alternatives to the planar surface code. Unlike planar surface code, QLDPC codes are space-efficient, but this efficiency comes at the price of denser connectivity beyond the degree-4 connectivity required for the planar surface code~\cite{tillich2013hgpcodes, breuckmann2021balancedproductcodes, delfosse2013hyperboliccodes, breuckmann2017hyperbolicsurfacecodes, higgott2023hyperbolicfloquetcodes}.
Furthermore, fault-tolerant syndrome extraction and decoding are relatively unexplored, especially under realistic noise models. Figure~\ref{fig:intro}(a) presents the two extremes of current error correction proposals. Planar surface codes have been demonstrated as they have simple connectivity requirements~\cite{krinner2022distance3demonstration, google2023suppressing}. However, it is hard to scale because of its poor rate. Contrarily, QLDPC codes are efficient but demand connectivity beyond degree-4. In this paper, we tackle the problems of \textit{constructing architectures for QLDPC codes}, \textit{fault-tolerant syndrome extraction}, and \textit{decoding}. We focus on two understudied QLDPC code families: \textit{hyperbolic surface codes}~\cite{breuckmann2017hyperbolicsurfacecodes, breuckmann2016constructionsofhyperbolicsurfacecodes}, and \textit{hyperbolic color codes}~\cite{delfosse2013hyperboliccodes}.

\vspace{0.05in}
\noindent
\textbf{Reducing Connectivity Requirements:} 
Figure~\ref{fig:intro}(b) compares the connectivity demands of the planar surface code to that of a hyperbolic surface code. Locally, the planar surface code requires degree-4 connectivity, as each parity qubit is connected to four data qubits and vice versa. In contrast, the hyperbolic surface code requires degree-5 connectivity, where parity qubits are connected to five data qubits. This na\"{i}ve architecture for the hyperbolic surface code would have two issues. The first issue is clear: the degree-5 connectivity would be hard to fabricate. The less obvious problem is that even if such an architecture could be fabricated, it might not support fault-tolerant syndrome extraction, potentially yielding poor error rates in practice. Our goal is to construct sparse architectures that also support fault-tolerant syndrome extraction.

Prior work has proposed using \textit{flag} qubits to reduce connectivity requirements~\cite{chao2018flagqubits, lao2020flagbridgequbits, liou2023flaglookuptables, baireuther2019nndecwithflags, chamberland2020heavycodes, chamberland2020colorcodeswithflags}. A flag qubit, shown in Figure~\ref{fig:intro}(c), has two purposes during syndrome extraction. \textit{First}, the flag qubit reduces connectivity: the circuit shown in Figure~\ref{fig:intro}(c) entangles two data qubits with a parity qubit by using a flag qubit as an intermediary. \textit{Second}, measuring the flag qubit can detect errors that harm the effective distance. These flag measurements form a \textit{flag syndrome} that a decoder must correctly leverage to correct errors accurately. Consequently, \textit{overusing} flag qubits will overwhelm the decoder with unnecessary information and increase decoding complexity. Ideally, an architecture should use as few flag qubits as possible while supporting fault-tolerant syndrome extraction.

To this end, we propose \textit{Flag-Proxy Networks (FPNs)}. FPNs, whose high-level layout is shown in Figure~\ref{fig:intro}(d), primarily use flag qubits to protect data qubits from errors that harm the effective distance while reducing connectivity demands. To avoid overusing flag qubits, further reductions in connectivity can be achieved by introducing ``proxy qubits," which need not be measured. To reduce the physical overheads of FPNs, we propose \textit{flag sharing}, which merges flag qubits common to the same data qubits.

\vspace{0.05in}
\noindent
\textbf{Syndrome Extraction Scheduling:} While FPNs provide an architecture to realize an error-correcting code, we must execute a syndrome extraction circuit to retrieve a syndrome and detect errors. Constructing valid and low-depth syndrome extraction schedules is an NP-hard problem~\cite{geher2024tanglingschedules, beverland2021costofuniversality, conrad2018bringscodecircuitlevel}. For codes with \textit{translation invariance}, such as the planar surface code~\cite{beverland2021costofuniversality}, a valid schedule for a single check can be reused for other checks in the code, thus significantly simplifying the problem. However, QLDPC codes do not necessarily have translation invariance. We present a \textit{greedy scheduling algorithm} for syndrome extraction scheduling for such codes. Our greedy algorithm uses a solver to schedule checks in isolation and imposes constraints on each check's schedule given already-scheduled checks.

\begin{figure*}[!htb]
    \centering
    \includegraphics[width=0.95\textwidth]{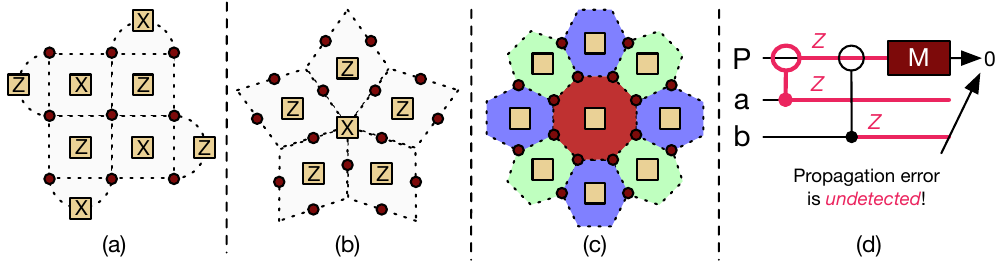}
    \caption{(a)~Example of planar surface code. (b)~Example of the local structure of a $\{4, 5\}$ hyperbolic surface code. Each edge corresponds to a data qubit, each face corresponds to an $X$ check, and each vertex corresponds to a $Z$ check. (c)~Example of a $\{4, 6\}$ hyperbolic color code. Each vertex corresponds to a data qubit, and each face (plaquette) corresponds to both a $Z$ and $X$ check. (c)~Example of an undetected propagation error caused by a CNOT error.}
    \label{fig:background}
\end{figure*}

\noindent
\textbf{Decoding with Flag Qubits:} As FPNs leverage flag qubits to detect errors during syndrome extraction, a decoder must leverage the additional \textit{flag syndrome} bits to identify errors correctly. In our analysis of error patterns on the hyperbolic codes, we find that errors that flip the \textit{same syndrome bits} but \textit{different flag bits} may correspond to entirely different data qubit errors. To handle this problem, we propose assigning flagged and unflagged errors into \textit{equivalence classes}. These equivalence classes contain error events that (1)~flip the same syndrome bits but (2)~flip different flag bits. During decoding, only one error event is considered from each equivalence class, given the flag syndrome. Finally, to evaluate our flag protocol, we propose two decoders that leverage this protocol to decode both flavors of hyperbolic codes accurately.

\vspace{0.05in}
\noindent
In summary, our paper makes the following contributions:
\begin{enumerate}
[leftmargin=0.4cm,itemindent=0.4cm,labelwidth=\itemindent,labelsep=0.05cm, align=left, itemsep=0.1cm, listparindent=0.5cm, topsep=0.1cm]
    \item We propose \textit{Flag-Proxy Networks}, a generalized architecture for quantum error correcting codes. Flag-Proxy Networks reduce connectivity demands by using flag and proxy qubits while enabling fault-tolerant syndrome extraction.
    \item We propose a \textit{greedy scheduling algorithm} applicable to any quantum code.
    \item We propose a flag syndrome protocol that organizes errors into equivalence classes. We further propose two decoders that leverage this protocol.
\end{enumerate}
Our evaluations demonstrate that FPNs of the hyperbolic surface and color codes are, respectively, $2.9\times$ and $5.5\times$ more efficient than the $d = 5$ planar surface code ($49$ physical qubits per logical qubit) while having comparable connectivity and error rates. This benefit only increases with larger distances. 
\section{Background}

\subsection{Characterizing Quantum Codes}
    Quantum error-correcting codes are often characterized by three properties: (1)~the number of data qubits in a logical block, (2)~the number of logical qubits in a logical block, and (3)~the code distance. The \textit{parameters} of a code are written as $[[n, k, d_X, d_Z]]$, where $n$ is the number of data qubits, $k$ is the number of logical qubits, $d_X$ is the code distance for $X$ errors, and $d_Z$ is the code distance for $Z$ errors. If $d_X = d_Z = d$, then the code may be written as $[[n,k,d]]$. Furthermore, $n-k$ checks must be measured to detect errors.

    Two other characteristics of quantum codes cannot be characterized entirely by $n$, $k$, and $d$. The first such characteristic is \textit{check weight}, which is the number of data qubits involved in the check. If a check has weight $\delta$, then the check's corresponding parity qubit must be connected to $\delta$ data qubits. The second characteristic is the total number of physical qubits $N$ required to implement a {\em logical block}, which depends on the underlying hardware and code. Standard implementations of the planar surface code on superconducting qubits require $N = 2n-1$ physical qubits. As $N \neq n$, we use two metrics to quantify a code's overheads: the \textit{ideal rate}, defined as $R_\mathrm{ideal} = k/n$, and the \textit{effective rate}, defined as $R_\mathrm{eff} = k/N$.

\subsection{Planar Surface Codes}
    First, we review the basics of the planar surface code. Each data qubit on the planar surface code is protected by at most two $Z$ checks and two $X$ checks, which correct $X$ and $Z$ errors and are sufficient to correct an arbitrary error on the data qubit. The most common implementation of the planar surface code is the \textit{rotated surface code}~\cite{horsman2012surfacecodelatticesurgery, tomita2014surfacecodewithrealisticnoise}, which is shown in Figure~\ref{fig:background}(a). This implementation of the planar surface code has parameters $[[d^2, 1, d]]$ and requires degree-4 connectivity. Secondly, fault-tolerant syndrome extraction for the rotated surface code can be implemented through proper CNOT ordering~\cite{tomita2014surfacecodewithrealisticnoise}. Finally, the rotated surface code can be decoded through \textit{Minimum-Weight Perfect Matching (MWPM)} decoding, which is fast and readily available. For these reasons, the planar surface code remains the predominant error-correcting code for superconducting systems. 

\subsection{A Brief Primer on QLDPC Codes}
    The pitfall of the planar surface code is that $R_\mathrm{ideal} = 1/d^2$, so error-corrected quantum computers using the planar surface code will require millions of physical qubits to support most applications~\cite{gidney2021rsafactorization, reiher2017nitrogenfixation, litinski2019gameofsurfacecodes}. Given the significant engineering hurdles necessary to support millions of qubits, the planar surface code is not an ideal candidate for error correction at scale.
    
    Instead of encoding a single logical qubit with a code, we want to encode multiple logical qubits. To do so, we must use codes with far fewer parity checks than data qubits: if a code has $n$ data qubits and $x$ parity checks, it will encode $k=n-x$ logical qubits. Figure~\ref{fig:qldpcconn}(a) shows the setup of the surface code, where parity checks detect errors on four data qubits. If checks were more complex and could detect errors across more data qubits, as in Figure~\ref{fig:qldpcconn}(b), then we could use fewer parity checks. This is the fundamental idea behind \textit{Quantum Low-Density Parity Check (QLDPC)} codes, which use far fewer checks than the surface code to detect errors on data qubits, thus yielding more logical qubits. Unfortunately, this comes at the cost of denser connectivity: \textit{denser} codes generally encode \textit{more} logical qubits.
    \color{black}
    
    \begin{figure}[!htb]
        \centering
        \includegraphics[width=0.8\columnwidth]{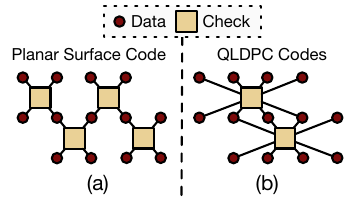}
        \caption{Parity checks for (a)~the planar surface code, and (b)~QLDPC codes. For illustration purposes only.}
        \label{fig:qldpcconn}
    \end{figure}

\subsection{Code Families for QLDPC Codes}
We note two challenges with realizing QLDPC codes on superconducting quantum computers. \textit{First}, QLDPC codes have high check-weights due to encoding multiple logical qubits. \textit{Second}, most QLDPC codes are understudied under realistic noise models; thus, our understanding of fault-tolerant syndrome extraction and decoding is limited for most codes. This section discusses these problems for two examples of QLDPC codes: \textit{hyperbolic surface codes} and \textit{hyperbolic color codes}. In this paper, we use ``hyperbolic codes" to collectively refer to hyperbolic surface and color codes.

    \subsubsection{Hyperbolic Surface Codes}
        Hyperbolic surface codes are created from geometric tilings parameterized by two integers $r$ and $s$ with the relationship in Equation~\eqref{eqn:hyperboliccondition} ~\cite{breuckmann2017hyperbolicsurfacecodes, breuckmann2016constructionsofhyperbolicsurfacecodes, breuckmann2022bringscodeandtraversalgates}.
        These $\{r, s\}$ tilings have the property that $s$ $r$-gons meet at a point. Each face corresponds to a $Z$ check, each point corresponds to an $X$ check, and each edge between two faces corresponds to a data qubit. Figure~\ref{fig:background}(b) shows an example of a $\{4, 5\}$ hyperbolic surface code.
        \begin{equation}
            \label{eqn:hyperboliccondition}
            \frac{1}{r} + \frac{1}{s} < \frac{1}{2}
        \end{equation}

    \subsubsection{Hyperbolic Color Codes}
        Hyperbolic color codes are constructed by selecting $r$ and $s$ with the same relationship as in Equation~\eqref{eqn:hyperboliccondition} with the additional constraint that $s$ is even~\cite{delfosse2013hyperboliccodes}. However, unlike hyperbolic surface codes, hyperbolic color codes are constructed by creating a tiling with three types of \textit{plaquettes}: \textit{red} plaquettes with $2r$ vertices and \textit{green} and \textit{blue} plaquettes with $s$ vertices. Each vertex in the tiling corresponds to a data qubit and is incident to one plaquette of each color, and each plaquette corresponds to an $X$ and $Z$ check. Figure~\ref{fig:background}(c) shows an example of a $\{4, 6\}$ hyperbolic color code.

    \subsubsection{Rate of Hyperbolic Codes}
        Each pair $\{r, s\}$ corresponds to a different \textit{subfamily} of hyperbolic codes, and each subfamily has a minimum ideal rate given by Equation~\eqref{eqn:hyperbolicrate}. Consequently, larger hyperbolic codes encode more logical qubits. In contrast, the planar surface code always encodes a single logical qubit.
        \begin{equation}
            \label{eqn:hyperbolicrate}
            R_\mathrm{ideal} \geq 1 - \frac{2}{r} - \frac{2}{s}
        \end{equation}

    \subsection{The Challenge of Connectivity}
        We note the following challenges with hyperbolic codes that are solved problems for the planar surface code:
        \begin{enumerate}
        [leftmargin=0.4cm,itemindent=0.4cm,labelwidth=\itemindent,labelsep=0.05cm, align=left, itemsep=0.1cm, listparindent=0.5cm, topsep=0.1cm]
            \item Hyperbolic surface codes require degree-$r$ and degree-$s$ connectivity for $Z$ and $X$ checks, respectively, whereas hyperbolic color codes require degree-$2r$ and degree-$s$ connectivity for each plaquette. In contrast, the planar surface code requires degree-4 connectivity.
            \item Hyperbolic codes have mostly not been studied under circuit-level noise beyond small examples~\cite{conrad2018bringscodecircuitlevel}. Thus, fault-tolerant syndrome extraction and decoding remain open research areas for most of the hyperbolic codes. In contrast, all three areas are well-studied for the planar surface code~\cite{tomita2014surfacecodewithrealisticnoise, higgott2022pymatching, higgott2023sparseblossom}.
        \end{enumerate}
        Given these challenges, we believe that the hyperbolic codes are representative of some obstacles blocking the realization of QLDPC codes on superconducting systems. In this paper, we consider codes with $n \leq 3000$ for four subfamilies of hyperbolic surface and color codes. Tables~\ref{tab:hysctbl} and \ref{tab:hycctbl} in the Appendix list the hyperbolic codes considered in this paper.

\subsection{The Challenge of Syndrome Extraction}
    In principle, a distance $d$ quantum code can correct $\lfloor (d-1)/2 \rfloor$ errors. However, operation errors in a syndrome extraction circuit can limit the effective distance ($d_\mathrm{eff}$) of a quantum code, limiting the correction capability of the code to only $\lfloor (d_\mathrm{eff}-1)/2 \rfloor$ errors. \color{black} Ideally, we want $d_\mathrm{eff} = d$: such a syndrome extraction circuit is considered \textit{fault-tolerant}.
    
    Consider a syndrome extraction circuit where a parity qubit interacts directly with each data qubit, as in Figure~\ref{fig:background}(c) for a $Z$ parity qubit $P$ interacting with two data qubits, $a$ and $b$. First, $CNOT(a, P)$ fails and causes a $Z$ error on both $a$ and $P$. While the $Z$ error on $P$ does not affect the parity outcome during measurement, it will propagate to $b$ through $CNOT(b, P)$; note this occurs regardless of whether the CNOT fails. By the end of the syndrome extraction round, both $a$ and $b$ have $Z$ errors. Note that this \textit{two-qubit data error} has occurred from a \textit{single fault} in the syndrome extraction circuit. We call such errors \textit{propagation errors}. 

    \vspace{0.05in}
    \noindent
    \textbf{Impact of propagation errors:} We explain how propagation errors impact $d_\mathrm{eff}$ at a high level. First, consider a $d=3$ code, which is only guaranteed to protect against one data error. A $d=3$ code cannot handle a propagation error because a single propagation error affects multiple qubits. However, the error stemmed from a single CNOT error; hence, as the $d = 3$ code cannot handle a single operation error, so $d_\mathrm{eff} = 2$. 
    
    \vspace{0.05in}
    \noindent
    \textbf{Fault-Tolerance in the Planar Surface Code:} Note that syndrome extraction in standard implementations of the planar surface code can tolerate propagation errors by reordering CNOTs~\cite{tomita2014surfacecodewithrealisticnoise}. Such a property is due to the structure of the planar surface code. While other codes can achieve fault-tolerance by reordering CNOTs~\cite{chao2018flagqubits, manes2023distancepreservinghgp, conrad2018bringscodecircuitlevel}, we explore leveraging flag qubits as a general strategy for fault-tolerance.

\subsection{The Challenge of Decoding}
    As a decoder must correct errors encountered during program execution, its performance is closely tied to the syndrome extraction circuitry. If the syndrome extraction circuit is not fault-tolerant, the decoder cannot correct more than $(d_\mathrm{eff}-1)/2$ errors. Similarly, an ineffective decoder may harm the effective distance even if the syndrome extraction circuit is fault-tolerant. For most QLDPC codes, this inter-relatedness between syndrome extraction and decoding has not been considered, as very little research considers the impact of circuit-level noise.

\subsection{Goal}
    QLDPC codes promise efficient error-corrected quantum computers but have challenges in (a)~dense connectivity, (b)~fault-tolerant syndrome extraction, and (c)~accurate decoding. In this paper, we develop a general, low-connectivity architecture for QLDPC codes while tackling the interrelated problems of fault-tolerant syndrome extraction and decoding.

\section{Evaluation Methodology}
\subsection{Error Model}
    \label{sec:errorandtiming}
    This paper considers a circuit-level error model, which best reflects errors found in real systems. Our error model contains the following errors for a physical error rate $p$.
    \begin{enumerate}[leftmargin=0.4cm,itemindent=0.4cm,labelwidth=\itemindent,labelsep=0.05cm, align=left, itemsep=0.1cm, listparindent=0.5cm, topsep=0.1cm]
        \item Decoherence and dephasing errors at the beginning of a syndrome extraction round. We assign each qubit a $T_1 = (1/p) \,\mu$s and $T_2 = 0.5T_1$ to model decoherence and idling error. Then, given a syndrome extraction latency $t$, $X$, $Y$, and $Z$ errors are injected with probability $p_X$, $p_Y$, and $p_Z$ according to the Pauli twirling approximation as described in Equations~\eqref{eqn:pxytwirling} and~\eqref{eqn:pztwirling}~\cite{tomita2014surfacecodewithrealisticnoise}. Specific operation latencies are listed below.
        \begin{equation}
            \label{eqn:pxytwirling}
            p_X = p_Y = \frac{1 - e^{-t/T_1}}{4}
        \end{equation}
        \begin{equation}
            \label{eqn:pztwirling}
            p_Z = \frac{1 - 2e^{-t/T_2} + e^{-t/T_1}}{4}
        \end{equation}
            
        \item Single-qubit gates cause random depolarizing errors at a rate of $0.1p$ and have a latency of $30$ns.
        \item Two-qubit gates cause random two-qubit depolarizing errors at a rate of $p$ and have a latency of $40$ns.
        \item Measurements return incorrect outcomes at a rate of $p$ and have a latency of $800$ns.
        \item Resets fail at a rate of $0.1p$ and have a latency of $30$ns.
        \item Idling errors occur during each two-qubit gate on qubits unused during the gate at a rate of $0.1p$.
    \end{enumerate}
    Unlike prior work, which fixes decoherence and dephasing errors to occur with probability $p$, using $T_1$ and $T_2$ times to model decoherence and dephasing errors penalizes longer syndrome extraction latencies. For instance, $2\times$ higher syndrome extraction latency results in $2\times$ higher $T_1$ and $T_2$ errors. Longer circuits also incur more idling errors. 
    Finally, we perform our simulations using Google's \textit{Stim} simulator~\cite{gidney2021stim}.

\subsection{Evaluating Architectural Overheads}
    We consider effective rate $R_\mathrm{eff} = k/N$, where $k$ is the number of logical qubits in a logical block, and $N$ is the total number of physical qubits required to realize the block.

\subsection{Evaluating Block Error Rate}
    In this paper, we execute memory experiments to evaluate a code's \textit{block error rate} ($BER$). A single memory experiment tests the code's capability to preserve an initial state (either $\ket{00\cdots 0}$ or $\ket{++\cdots +}$) in the presence of errors over $d$ syndrome extraction rounds. Once the $d$ rounds are finished, the resulting syndrome is given to a decoder, which tries to correct any logical qubits. If any logical qubits have an error after decoding, an error has occurred.
    
    Thousands of memory experiment trials are executed to estimate the $BER$, defined in Equation~\eqref{eqn:ber}. We use the \textit{normalized block error rate} $BER_\mathrm{norm} = BER/k$ to compare codes of different block sizes.
    \begin{equation}
        \label{eqn:ber}
        BER = \frac{\text{number of errors on any logical qubit}}{\text{number of fault-injection trials}}
    \end{equation}
\vspace{0.1in}
\section{Flag-Proxy Networks}
This section presents \textit{Flag-Proxy Networks}, a microarchitectural paradigm that makes QLDPC codes amenable to superconducting architectures.

\subsection{Flag Qubits}
    Reducing connectivity requires introducing additional qubits to enable interactions between non-adjacent qubits. One such approach involves using \textit{flag qubits}\footnote{Flag qubits are also known as ``(flag-)bridge" qubits in prior work~\cite{lao2020flagbridgequbits, wu2022surfstitch, yin2023codestitch} to emphasize that flag qubits reduce connectivity demands.}~\cite{chao2018flagqubits, lao2020flagbridgequbits, wu2022surfstitch, yin2023codestitch}. Flag qubits not only reduce connectivity, but with correct use, they can detect propagation errors. Figure~\ref{fig:flagexample} presents the example of a flag qubit $F$ used to entangle data qubits $a$ and $b$ with a parity qubit $P$. Here, a propagation error on $a$ and $b$ affects the phase of $F$. At the end of the syndrome extraction circuit, $F$ is measured, and this measurement yields one, indicating a propagation error has occurred.

    \begin{figure}[!htb]
        \centering
        \includegraphics[width=\columnwidth]{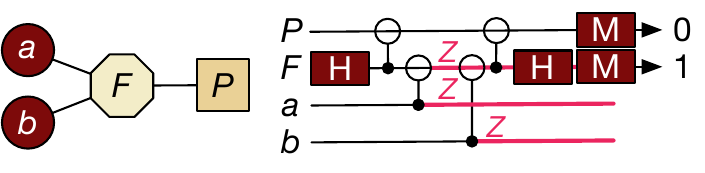}
        \caption{Left: local connectivity required by a flag qubit. Right: corresponding syndrome extraction with a flag qubit $F$, which detects a propagation error.}
        \label{fig:flagexample}
    \end{figure}

    \vspace{0.1in}
    \noindent
    \textbf{Fault-Tolerance with Flag Qubits:} While flag qubits can detect propagation errors, simply introducing flag qubits is insufficient for fault-tolerance. Indeed, the flag measurements form a secondary syndrome known as the \textit{flag syndrome}. The decoder must leverage this flag syndrome to account for propagation errors during syndrome extraction. Furthermore, flag measurement errors cannot be detected, unlike parity measurements, as a propagation error will not repeat between rounds. Thus, an accurate decoder must account for flag measurement errors during decoding. We observe that while prior work has used flag qubits to reduce connectivity demands, they have not considered how to leverage flag qubits more generally during decoding~\cite{lao2020flagbridgequbits, wu2022surfstitch, yin2023codestitch}.

    \vspace{0.1in}
    \noindent
    \textbf{Flag Overuse:} 
    As a decoder must use flag qubits, \textit{overusing} flag qubits where they are unnecessary may overburden the decoder with useless information. An example of such a situation is shown in Figure~\ref{fig:flagoveruse}, which shows a syndrome extraction circuit with three flags $F$, $G$, and $H$. Here, while $H$ detects a propagation error on $F$ and $G$, $F$ and $G$ both detect the same propagation error. Thus, measuring $H$ is unnecessary as $F$ and $G$ detect the propagation error.

    \begin{figure}[!htb]
        \centering
        \includegraphics[width=\columnwidth]{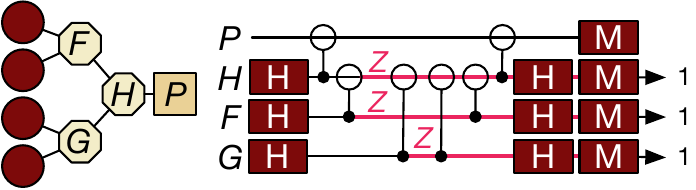}
        \caption{An example of a flag $H$ which provides redundant information about a propagation error detected by flags $F$ and $G$.}
        \label{fig:flagoveruse}
    \end{figure}
    
\subsection{Proxy Qubits}
    As an alternative to flag qubits, we propose \textit{proxy qubits} as a secondary mechanism for sparsifying connectivity. Unlike flag qubits, proxy qubits need not be measured nor entangled with parity qubits at the start of a syndrome extraction round. Thus, they avoid the same overuse problem seen with flag qubits. Figure~\ref{fig:proxyexample} depicts the example of a proxy qubit \textit{x} used to entangle data qubits \textit{a} and \textit{b} with parity qubit $P$. Here, the proxy qubit $x$, initialized in $\ket{0}$, is entangled with $a$ to form a GHZ state after $CNOT(a, x)$. Then, operation $CNOT(x, P)$ effectively performs the CNOT between $a$ and $P$. Finally, $CNOT(a, x)$ undoes the GHZ state, ideally returning $x$ to $\ket{0}$. This same procedure is repeated for qubit $b$.

    \begin{figure}[!htb]
        \centering
        \includegraphics[width=0.85\columnwidth]{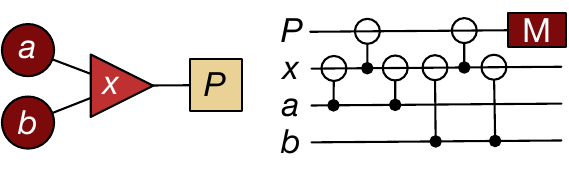}
        \caption{Proxy $x$ is used to entangle data qubits $a$ and $b$ with parity qubit $P$.}
        \label{fig:proxyexample}
    \end{figure}

    Nevertheless, proxy qubits are susceptible to certain errors during syndrome extraction. We discuss how three such errors can be managed to avoid harming the block error rate.

    \vspace{0.1in}
    \noindent
    \textbf{Type 1 Errors:} These errors are $X$ ($Z$) errors when measuring a $Z$ ($X$) check. We find that these errors result in the proxy qubit returning to the $\ket{1}$ state instead of the $\ket{0}$ state, which results in measurement-like errors on the parity measurements. These errors do not reduce the effective distance. 
    
    \vspace{0.1in}
    \noindent
    \textbf{Type 2 Errors:} These errors result from improper CNOT orientation, namely $Z$ errors while measuring an $X$ check. Figure~\ref{fig:orientationexample}(a) and Figure~\ref{fig:orientationexample}(b) present two ways of entangling a data qubit $a$ and a parity qubit $P$ via a proxy $x$. Figure~\ref{fig:orientationexample}(a)'s method is more erroneous as it causes more measurement-like errors when measuring $P$. These errors stem from three sources:
    \begin{enumerate}[leftmargin=0.4cm,itemindent=0.5cm,labelwidth=\itemindent,labelsep=0cm, align=left, itemsep=0.0cm, listparindent=0.5cm, topsep=0.15cm]
        \item A $Y$/$Z$ error on $x$ or $P$ in the first CNOT (about $p/2$).
        \item A $Y$/$Z$ error on $x$ in the second CNOT ($p/4$).
        \item A $Y$/$Z$ error on $P$ in the last CNOT ($p/4$).
    \end{enumerate}
    In total, the error probability is $p$. In contrast, Figure~\ref{fig:orientationexample}(b)'s circuit only has an error probability $p/2$, as only $Y$/$Z$ errors in the first two CNOTs can cause a measurement-like error. Our studies also confirm that Figure~\ref{fig:orientationexample}(b)'s circuit yields lower error rates.

    \begin{figure}[!htb]
        \centering
        \includegraphics[width=0.95\columnwidth]{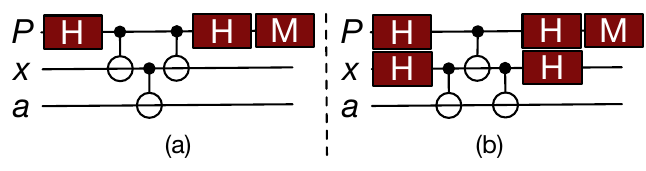}
        \caption{Examples of two possible CNOT orders to entangle $a$ and $P$ through a proxy $x$.}
        \label{fig:orientationexample}
        \vspace{-0.1in}
    \end{figure}

\begin{figure*}[!t]
    \centering
    \includegraphics[width=\textwidth]{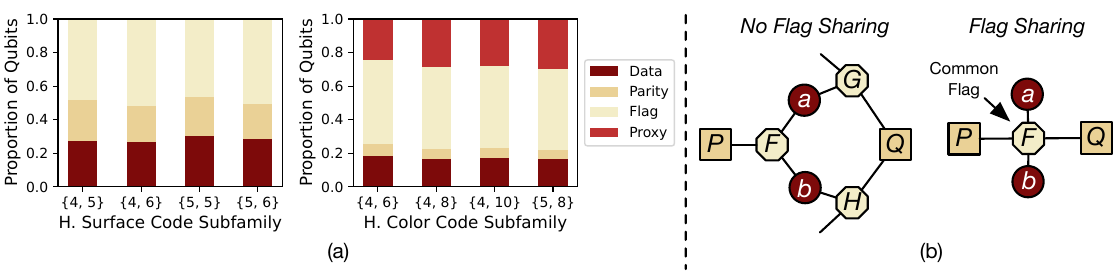}
    \caption{(a)~Qubit overheads by type (data, parity, etc.) for subfamilies of hyperbolic surface and color codes. (b)~Example of flag sharing applied to data qubits $a$ and $b$, which have common checks $P$ and $Q$. Flag sharing reduces overheads and connectivity demands.}
    \label{fig:flagsharing}
\end{figure*}

    \vspace{0.1in}
    \noindent
    \textbf{Type 3 Errors:} These are propagation errors that result from the misapplication of proxy qubits, as in Figure~\ref{fig:badproxy}. Here, data qubits $a$ and $b$ are simultaneously entangled to parity qubit $P$ via proxy qubit $x$. When a $Z$ error occurs on $x$, it propagates to $a$ and $b$. To avoid such errors, $a$ and $b$ must be entangled to $P$ separately. Note that such errors only occur when $a$ and $b$ are both data qubits. If both qubits are flags, they can be entangled simultaneously with $P$, as any resulting propagation error is detectable by measuring the flag qubits.

    \begin{figure}[!htb]
        \centering
        \includegraphics[width=0.45\columnwidth]{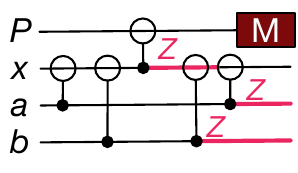}
        \caption{Misusing proxy $x$ results in a propagation error onto data qubits $a$ and $b$.}
        \vspace{-0.1in}
        \label{fig:badproxy}
    \end{figure}

    \vspace{0.1in}
    \noindent
    \textbf{Measuring Proxy Qubits?} We found that measuring proxy qubits (like flag qubits) to detect errors results in rather complex syndromes. We leave this avenue for future research.

\subsection{Flag-Proxy Networks}
    Flag and proxy qubits offer two methods of reducing connectivity requirements. While one or the other can be used indiscriminately, we note that both approaches have pitfalls. Only using flag qubits will result in flag overuse, which can harm the decoder, whereas only using proxy qubits will not guarantee fault-tolerant syndrome extraction. Ideally, we want an architecture that supports fault-tolerant syndrome extraction without overusing flag qubits.

    Our key insight towards this goal is as follows: \textit{given a minimal set of flag qubits that protect data qubits from propagation errors, proxy qubits can further reduce connectivity without harming the effective distance}. We present this argument formally in Theorem~\ref{th:faulttolerance}, whose proof is available in Appendix~\ref{app:sec:proofoftheorem}. Nevertheless, we leverage this insight to design \textit{Flag-Proxy Networks (FPNs)}. FPNs meet all our criteria for a good architecture: (1)~its connectivity can be made arbitrarily low with flag and proxy qubits, (2)~it supports fault-tolerant syndrome extraction by leveraging flag qubits, and (3)~it avoids flag overuse by using proxy qubits.

    In general, computing a minimal set of flag qubits is difficult and depends on syndrome extraction, a code's logical operators, and the decoder~\cite{chao2018flagqubits, chamberland2020colorcodeswithflags}. For simplicity, we set up the flag layer as in Figure~\ref{fig:flagsetup}, where $\delta/2$ flags detect propagation errors in a weight-$\delta$ check such that each flag is assigned to a pair of data qubits. This setup is fault-tolerant, as any propagation error can be detected by one or more flags and is amenable to the decoders we consider in this paper.
    
    \begin{thm}
    \label{th:faulttolerance}
        Suppose that a Flag-Proxy Network without proxies is fault-tolerant. The same network with proxies is also fault-tolerant.
    \end{thm}

    \begin{figure}[!htb]
        \centering
        \includegraphics[width=0.4\columnwidth]{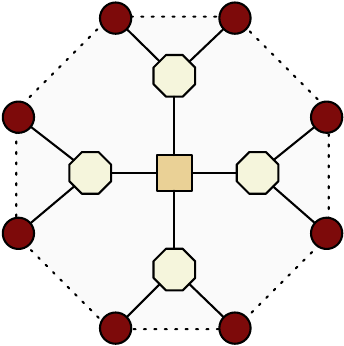}
        \caption{Flag setup for a weight-8 check. Note that this setup may need proxies to meet connectivity constraints.}
        \vspace{-0.1in}
        \label{fig:flagsetup}
    \end{figure}
    
\subsection{Constructing Flag-Proxy Networks}
We briefly discuss how to construct FPNs given a quantum code. To begin with, start with a na\"{i}ve architecture that connects data qubits to parity qubits. While this architecture may violate connectivity constraints, we can introduce flag and proxy qubits to alleviate the connectivity demands. Flag qubits should be introduced according to a fault-tolerant flag protocol. For simplicity, we use the flag protocol in Figure~\ref{fig:flagsetup}, which uses many flags but is guaranteed to be fault-tolerant. After introducing flags, any high-degree qubits can be reduced to lower-degree qubits through proxy qubits, as in Figure~\ref{fig:proxyd6ex}.

\begin{figure}[!htb]
    \centering
    \includegraphics[width=0.6\columnwidth]{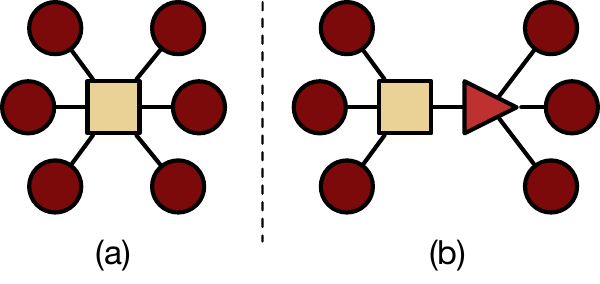}
    \vspace{-0.1in}
    \caption{(a)~A degree-6 qubit transformed into (b)~a degree-4 qubit by adding a proxy qubit.}
    \label{fig:proxyd6ex}
    \vspace{-0.1in}
\end{figure}

\vspace{0.05in}
\noindent
\textbf{How many flag and proxy qubits?} The number of flag and proxy qubits depends on the underlying quantum code and flag protocol. If a dense code is used with a flag protocol that uses few flags, then many proxies may be required to maintain connectivity constraints. With the flag protocol used in this paper, the hyperbolic surface codes do not need proxy qubits as they have, at worst, degree-6 connectivity. In contrast, the hyperbolic color codes are very dense and need a few (at most three) proxy qubits to achieve degree-4 connectivity.

\subsection{Reducing Overheads with Flag Sharing}
    \label{sec:flagsharing}
    Figure~\ref{fig:flagsharing}(a) shows the average qubit composition within an FPN for different subfamilies of hyperbolic codes. Flag qubits make up \textit{almost half} of all qubits and are thus the most significant contributor to physical overheads. Ideally, we want to support fault-tolerant syndrome extraction without having such exorbitant overheads.

    To this end, we propose \textit{flag sharing} within FPNs, as shown in Figure~\ref{fig:flagsharing}(b). Our key insight here is that checks often share at least two data qubits; note that this is true for practically all error-correcting codes. Thus, we merge flag qubits between checks with common data qubits to reduce the number of flag qubits. To optimize overheads across the entire code, we pair data qubits together using \textit{maximum weight matching}, where the weight between a pair of qubits is the number of common checks. With this strategy, flag sharing reduces flag overheads by 10\%. Furthermore, flag sharing also removes the need for proxy qubits outside the {4,10} and {5,8} hyperbolic color codes subfamilies.

    \vspace{0.1in}
    \noindent
    \textbf{Results:}
    Figure~\ref{fig:effrate} further compares the effective rate of FPNs for hyperbolic codes with and without flag sharing. The effective rate for the standard implementation of a $d = 5$ planar surface code is marked for reference. Flag sharing improves the effective rate by $1.2\times$ and $2.4\times$ for hyperbolic surface and color codes, respectively. We also find that FPNs for the hyperbolic codes strictly outperform the $d = 5$ planar surface code, which has an effective rate of $1/49$. FPNs of the hyperbolic surface codes outperform the $d = 5$ planar surface code by $2.9\times$ on average and up to $4.6\times$. Concurrently, FPNs of the hyperbolic color codes outperform the $d = 5$ planar surface code by $5.5\times$ on average and up to $6.8\times$. These FPNs will only further outperform the planar surface code when considering larger code distances.

    \begin{figure}[!htb]
        \centering
        \includegraphics[width=0.95\columnwidth]{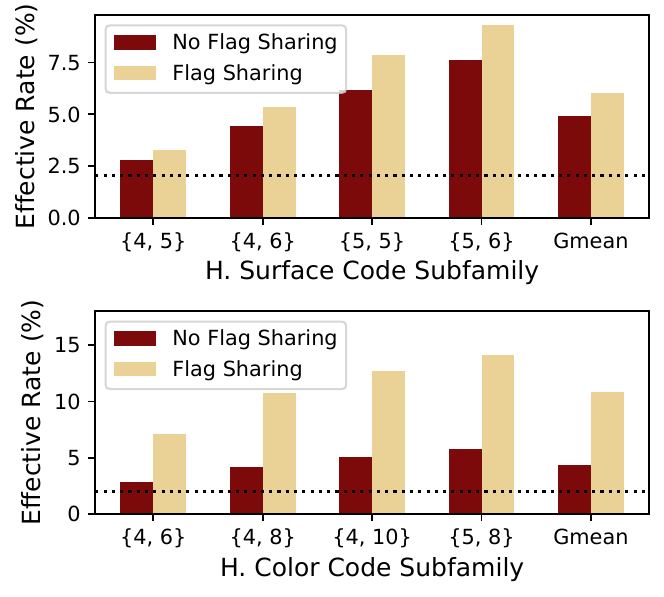}
        \caption{Effective rates for FPNs with and without flag sharing for hyperbolic codes. The effective rate of a $d = 5$ planar surface code ($1/49$) is marked for reference.}
        \label{fig:effrate}
    \end{figure}

    Table~\ref{tab:meanconn} compares the highest mean degree of an FPN (with flag sharing) in each subfamily to the standard implementations of the $d = 3, 5, 7$ planar surface codes. Note that the the maximum degree of each FPN is four, the same as the planar surface code. The lower connectivity of the hyperbolic codes is because flag sharing causes each data qubit to be connected to two flag qubits. In contrast, most data qubits in the surface code are connected to three or four parity qubits.

    \begin{table}[!htb]
        \centering
        \begin{center}
        \caption{Highest Mean Degree by Subfamily}
        \label{tab:meanconn}
        \begin{tabular}{|c|c||c|}
        \hline
        Family & Subfamily & Highest Mean Conn. \\
        \hline
        \hline
        \multirow{4}{*}{H. Surface Code} 
            & $\{4,5\}$ & 2.98  \\
        \cline{2-3}
            & $\{4,6\}$ & 2.94 \\
        \cline{2-3} 
            & $\{5,5\}$ & 3.12 \\
        \cline{2-3}
            & $\{5,6\}$ & 3.11 \\
        \hline
        \multirow{4}{*}{H. Color Code} 
            & $\{4,6\}$ & 2.80  \\
        \cline{2-3}
            & $\{4,8\}$ & 2.94 \\
        \cline{2-3} 
            & $\{4,10\}$ & 2.90 \\
        \cline{2-3}
            & $\{5,8\}$ & 2.93 \\
        \hline
        \hline
        \multirow{3}{*}{P. Surface Code}
            & $d=3$ & 2.82 \\
        \cline{2-3} 
            & $d=5$ & 3.26 \\
        \cline{2-3}
            & $d=7$ & 3.46 \\
        \hline
        \end{tabular}
        \end{center}
    \end{table}
    
\section{Syndrome Extraction}
While FPNs provide an architecture for a quantum code, detecting errors on the code requires executing a syndrome extraction circuit. In this section, we detail the challenges of syndrome extraction scheduling and present a general algorithm for syndrome extraction scheduling.

\subsection{Constraints of Syndrome Extraction}
    Functionally correct syndrome extraction schedules must abide by two constraints~\cite{beverland2021costofuniversality, geher2024tanglingschedules}. In this section, we use the notation $t_K(q)$ to refer to the timestep a qubit $q$ has a CNOT when measuring check $K$.
    \begin{enumerate}[leftmargin=0.4cm,itemindent=0.5cm,labelwidth=\itemindent,labelsep=0cm, align=left, itemsep=0.0cm, listparindent=0.5cm, topsep=0.15cm]
        \item[\textbf{Uniqueness: }] A data qubit $q$ can only perform one CNOT at a time. That is, $t_{K_i}(q) \neq t_{K_j}(q)$ where $K_i$ and $K_j$ are checks. Similarly, a parity qubit for $K_i$ can only perform one CNOT at a time: $t_{K_i}(q_1) \neq t_{K_i}(q_2)$ for $q_1, q_2 \in K_i$.
        \item[\textbf{Commutation: }] Let $K_X$ be an $X$ check and $K_Z$ be some $Z$ check. If $K_X$ and $K_Z$ have common qubits $\mathsf{Comm}$, then the relationship in Equation~\eqref{eqn:commrel} must hold.
        \begin{equation}
            \label{eqn:commrel}
            \prod_{q \in \mathsf{Comm}} \left( t_{K_X}(q) - t_{K_Z}(q)  \right) > 0
        \end{equation}
    \end{enumerate}
    Due to these two constraints, the worst-case latency for a syndrome extraction schedule is one where $X$ checks and $Z$ checks are measured disjointly; such a schedule has depth $\max ( \delta_X ) + \max ( \delta_Z )$ where $\delta_X$ and $\delta_Z$ are the $X$ and $Z$ check weights of the code. To the best of our knowledge, the only prior work on syndrome extraction scheduling is a coloring algorithm that guarantees this worst-case depth~\cite{tremblay2022thinplanarconnectvity}. We aim to achieve better-than-worst-case syndrome extraction depth to minimize decoherence, dephasing, and idling errors.

\subsection{Difficulty of Scheduling}
    Optimal low-depth scheduling is NP-Hard, and state-of-the-art solvers cannot compute optimal schedules beyond the smallest quantum codes. Indeed, QLDPC codes are rather large, and thus, solvers cannot used to schedule these codes. However, scheduling is relatively straightforward when considering planar codes, such as the planar surface code, as such codes are \textit{translation invariant}. Translation invariance implies that checks on the code are ``locally identical". For codes with translation invariance, optimal schedules can be obtained by computing a low-depth schedule for a handful of checks and then reusing these schedules for the entire code, as shown in Figure~\ref{fig:translationinvariance} for the planar surface code. Unfortunately, QLDPC codes are not necessarily translation invariant. For codes without translation invariance, we need an alternative method of computing syndrome extraction schedules.

    \begin{figure}[!htb]
        \centering
        \includegraphics[width=0.8\columnwidth]{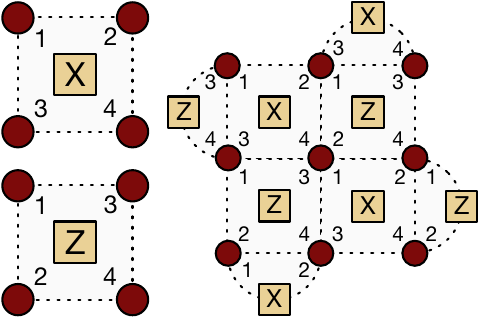}
        \caption{Translation invariance on the planar surface code, where the same $X$ and $Z$ schedules are reused for every such check.}
        \label{fig:translationinvariance}
    \end{figure}

\subsection{Reducing the Complexity of the Problem}
    Computing syndrome extraction schedules is not tractable for solvers for two reasons. \textit{First}, a solver will need at most $(n-k)\delta_\mathrm{max}$ ``time" variables ($t_K(q)$) to encode the scheduling problem, as $\delta$ variables are required for each weight-$\delta$ check. Thus, a code with about 100 qubits will require thousands of time variables. \textit{Second}, uniqueness and commutation constraints require additional ``auxiliary" variables and constraints to implement in practice. Equation~\eqref{eqn:uniquenessequivcon} presents the encoding of a uniqueness condition, where each condition requires two constraints and an additional auxiliary variable $x$. Commutation constraints, as in Equation~\eqref{eqn:commutationequivcon}, require $2|\mathsf{Comm}|+1$ constraints and $|\mathsf{Comm}|+1$ auxiliary variables. 

    \noindent
    \begin{footnotesize}
    \begin{equation}
        \label{eqn:uniquenessequivcon}
        t_{K_i}(q) \neq t_{K_j}(q) \rightarrow \begin{cases}
            t_{K_i}(q) - Mx \leq t_{K_j}(q) - 1 \\
            t_{K_i}(q) - M(1-x) \geq t_{K_j}(q) + 1 \\
            x \in \{0, 1\},\, M \gg 0
        \end{cases}
    \end{equation}
    \begin{equation}
        \label{eqn:commutationequivcon}
        \prod_{q \in \mathsf{Comm}} \left( t_{K_X}(q) - t_{K_Z}(q)  \right) > 0 \rightarrow \begin{cases}
            \sum_{q \in \mathsf{Comm}} x_q = 2y \\
            t_{K_X}(q) - t_{K_Z}(q) \leq Mx_q \\
            t_{K_Z}(q) - t_{K_X}(q) \leq M(1-x_q) \\
            x_q \in \{0, 1\},\, y > 0,\, M \gg 0
        \end{cases} 
    \end{equation}
    \end{footnotesize}
    \vspace{0.1in}
    
    Leveraging a solver to compute a schedule requires significantly reducing these variable and constraint overheads. Our insight towards this goal is to \textit{compute locally optimal schedules} instead of globally optimal ones. Locally optimal scheduling reduces variable and constraint overheads in two ways. \textit{First}, by scheduling a single weight-$\delta$ check, we only require $\delta$ time variables. \textit{Second}, uniqueness and commutation constraints only need to be considered for adjacent checks that share data qubits. Thus, the runtime of the solver instead depends on \textit{the size of the check} instead of the size of the code. We denote the runtime for a weight-$\delta$ check as $T(\delta)$.

\subsection{Greedy Scheduling Algorithm}
    We present a greedy algorithm for syndrome extraction scheduling, which leverages our prior insight, shown in Algorithm~\ref{alg:syndromeext}. The complexity of this algorithm, which schedules checks sequentially, is $O\left( (n-k)T(\delta_\mathrm{max}) \right)$ for an $[[n, k, d]]$ code. In practice, our algorithm is fast as $T(\delta_\mathrm{max}) \approx O(100\text{ms})$ for commercial solvers in the worst case.
    
    \begin{algorithm}[!htb]
        \caption{Greedy Scheduling Algorithm}
        \label{alg:syndromeext}
        \SetKwInput{KwInput}{Input}
        \SetKwInput{KwOutput}{Output}
        \DontPrintSemicolon

        \KwInput{Checks $K_1, \cdots, K_{n-k}$}
        \KwOutput{A CNOT Schedule}
        \hfill\;
        Let $\delta_\mathrm{max}$ be the maximum check weight.\;
        Create a table of scheduled CNOT times $T(K, q)$.\;
        \For{$1 \leq i \leq n-k$}{
            Create the following program for check $K_i$:
            \begin{align*}
                \text{min.} \quad& t_\mathrm{max} \\
                \text{s.t.} \quad& t_{K_i}(q_1) \neq t_{K_i}(q_2)
                    \quad\text{where } q_1, q_2 \in K_i
                    \\
                & t_{K_i}(q) \neq T(K_j, q) 
                    \quad\text{where } q \in \mathsf{Comm}(K_i, K_j) 
                    \\
                & \prod_{q \in \mathsf{Comm}(K_i, K_j)} (t_{K_i}(q) - T(K_j, q)) > 0 \\
                & t_\mathrm{max} \geq t_{K_i}(q) \\
                & 1 \leq t_{K_i}(q) \leq 2\delta_\mathrm{max}
            \end{align*}
            where $1 \leq j < i$.\;
            Use a solver to compute a solution to the program.\;
            Assign $T(K_i, q) := \mathsf{Result}(t_{K_i}(q))$ for all $q \in K_i$.\;
        }
        \KwRet $T(K, q)$ as a schedule of CNOTs.\;
    \end{algorithm}

\subsection{Validity of the Greedy Algorithm}
    To output a valid schedule, the greedy algorithm (Algorithm~\ref{alg:syndromeext}) must abide by global uniqueness and commutativity constraints. We discuss how the greedy algorithm does so while scheduling checks sequentially. Without loss of generality, suppose the greedy algorithm is currently scheduling a check $K_i$ and has already scheduled check $K_{i-1}$, and suppose $K_i$ and $K_{i-1}$ share two qubits, $a$ and $b$. Note that the steps below are repeated for any scheduled check $K_j$ that shares qubits with $K_i$.

    \subsubsection{Uniqueness} As $K_{i-1}$ has already been scheduled, we know that qubits $a$ and $b$ are already scheduled to perform a CNOT. Say these times are $t_a = T(K_{i-1},a)$ and $t_b = T(K_{i-1},b)$. Then, when scheduling check $K_i$, the greedy algorithm requires that qubits $a$ and $b$ are not scheduled at times $t_a$ and $t_b$, respectively. This ensures that qubits $a$ and $b$ perform only one CNOT at any given time, ensuring global uniqueness constraints are maintained.

    \subsubsection{Commutativity} As commutativity constraints need not be applied if $K_i$ and $K_{i-1}$ are both $X$ or both $Z$ checks, without loss of generality, assume $K_i$ is a $Z$ check and $K_{i-1}$ is an $X$ check. Like before, we know that qubits $a$ and $b$ are already scheduled for a CNOT for check $K_{i-1}$ at times $t_a$ and $t_b$. To ensure global commutativity constraints are met, the greedy algorithm directs the solver to enforce the constraint $(t_a' - t_a)(t_b' - t_b) > 0$, where $t_a', t_b'$ are solver variables.

\subsection{Performance of Greedy Algorithm}
    \label{sec:algperf}
    To evaluate the performance of the greedy algorithm, we first discuss the \textit{theoretically shortest circuit} for each code. 
    For a code with a maximum check weight of $\delta$, the theoretically shortest circuit will have $2$ $H$ gates, $\delta$ CNOT gates, and a measurement+reset gate. On the other hand, the \textit{longest possible circuit} will forego commutation constraints and schedule $X$ and $Z$ checks separately and thus will have $2$ $H$ gates, $\delta_X+\delta_Z$ CNOT gates, and a measurement+reset gate. Under our timing model from Section~\ref{sec:errorandtiming}, the shortest possible circuit and longest possible circuit will have latencies of $(890 + 40\delta)\mathrm{ns}$ and $(890 + 40\delta_X + 40\delta_Z)\mathrm{ns}$. 
    
    Figure~\ref{fig:roundlat} compares the output syndrome extraction circuits from the greedy algorithm to that of the theoretically shortest and longest circuits. In all cases but for the $\{4,5\}$ hyperbolic surface codes, the mean latency observed is less than that of the theoretical longest latency. Furthermore, we find the greedy algorithm performs better for denser codes, as the difference between the theoretical shortest and longest latency is much larger for these codes. Nevertheless, our algorithm is the first to offer better-than-worst-case syndrome extraction latency, and we expect future work to improve upon our results.

    \begin{figure}[!htb]
        \centering
        \includegraphics[width=\columnwidth]{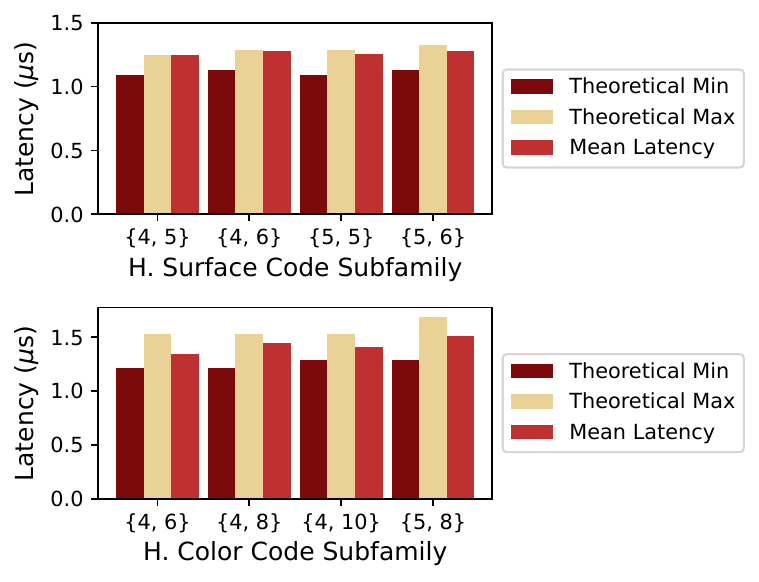}
        \caption{Syndrome extraction latencies for schedules computed by the greedy algorithm (Algorithm~\ref{alg:syndromeext}).}
        \label{fig:roundlat}
    \end{figure}

\subsection{Scheduling for FPNs}
    Algorithm~\ref{alg:syndromeext} schedules CNOTs between data and parity qubits and can be used outside of FPNs. However, by itself, Algorithm~\ref{alg:syndromeext} does not (1)~consider the layout dictated by an FPN and (2)~support fault-tolerant syndrome extraction with flag and proxy qubits. In this section, we discuss minor modifications to the algorithm to operate with FPNs.

    \subsubsection{Flag Qubit Modifications} 
        Flag qubits must be (1)~initialized, (2)~perform CNOTs with data qubits, and (3)~measured. During initialization and measurement, flag qubits perform CNOTs with the parity qubits. These CNOTs have no constraints beyond uniqueness and can be done greedily. When performing CNOTs with data qubits, flag qubits present an opportunity for parallelism during syndrome extraction. To leverage this parallelism, uniqueness constraints must be modified by replacing all parity qubit constraints with flag qubit constraints. If a flag qubit $F$ operates on qubits $q_1$ and $q_2$, then $t_F(q_1) \neq t_F(q_2)$ must hold. Note that if $q_1, q_2 \in K$ and $F$ is used in the syndrome extraction of $K$, then $t_F(q_1) = t_{K}(q_1)$ and $t_F(q_2) = t_{K}(q_2)$. Furthermore, if checks $K_i$ and $K_j$ share flag $F$, then the constraint $t_{K_i}(q) \neq T(K_j, q)$ must be replaced with $t_{K_i}(q) = T(K_j, q)$ as $q$ has already been entangled with $K_i$ through $F$.

    \subsubsection{Proxy Qubit Modifications}
        Proxy qubits do not require any modifications in Algorithm~\ref{alg:syndromeext} beyond handling CNOTs between two non-adjacent qubits. Here, we compute the shortest path between two non-adjacent qubits such that the interior edges of this path do not pass over data, parity, or flag qubits. Then, we perform CNOTs along this path.

    \subsubsection{Results} 
        We briefly compare the output syndrome extraction latencies for FPNs of the hyperbolic codes to that of a standard implementation of the planar surface code. Under our timing model in Section~\ref{sec:errorandtiming}, the planar surface code has a syndrome extraction latency of about $1\mu$s. In comparison, the hyperbolic surface and color codes have worst-case latencies of $2.3\mu$s and $3.4\mu$s. The longer latency of the hyperbolic surface code results from flag sharing, and the longer latency of the hyperbolic color codes result from $X$ and $Z$ checks being measured separately. Nevertheless, these latencies are comparable to that of the planar surface code.

\section{Decoding With Flag Qubits}
While FPNs realize error-correcting codes and provide fault-tolerant syndrome extraction circuits, correcting errors requires using a decoder to identify data qubit errors. In this section, we present a generalizable flag protocol applicable to FPNs. We further present modifications of \textit{Minimum-Weight Perfect Matching (MWPM)}~\cite{edmonds1965mwpm, higgott2023sparseblossom, higgott2022pymatching} and \textit{Restriction}~\cite{kubica2023matchingandlifting, chamberland2020colorcodeswithflags} decoders which leverage this flag protocol to decode the hyperbolic codes.

\begin{figure}[!htb]
    \centering
    \includegraphics[width=0.95\columnwidth]{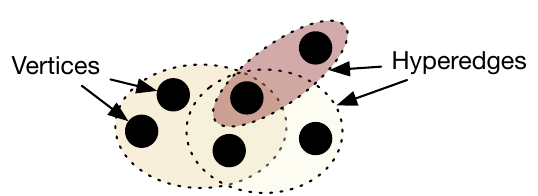}
    \caption{A hypergraph with three hyperedges. The primary difference between a graph and a hypergraph is that a hyperedge can connect $\geq 2$ vertices, whereas an edge can only connect $2$ vertices.}
    \label{fig:hypergraphex}
    \vspace{-0.2in}
\end{figure}

\subsection{Decoding Hypergraph}
    We represent all possible errors during syndrome extraction in a \textit{decoding hypergraph}. 
    Hypergraphs, shown in Figure~\ref{fig:hypergraphex}, extend graphs by using \textit{hyperedges}, which connect arbitrarily many vertices. The vertices of a decoding hypergraph correspond to syndrome bits, and the hyperedges correspond to error events. A hyperedge has the following properties:
    \begin{enumerate}[leftmargin=0.4cm,itemindent=0.5cm,labelwidth=\itemindent,labelsep=0cm, align=left, itemsep=0.0cm, listparindent=0.5cm]
        \item A set of syndrome bits $\sigma(e)$ flipped by the corresponding error event.
        \item A set of flag bits $f(e)$ flipped by the corresponding error event.
        \item An error probability $\pi(e)$.
        \item A set of affected Pauli frames $\lambda(e)$. Each Pauli frame corresponds to either an $X$ or $Z$ error on a logical qubit. These errors are tracked by the decoder in the software.
    \end{enumerate}
    Finally, we say a hyperedge is a \textit{flag hyperedge} if $|f(e)| > 0$ and otherwise call it a \textit{normal hyperedge}.

\subsection{Error Equivalence Classes}
    Given the flag syndrome from syndrome extraction, we must determine whether or not to use flag hyperedges. For hyperbolic codes, we observe the following patterns in flag hyperedges:
    \begin{enumerate}[leftmargin=0.4cm,itemindent=0.5cm,labelwidth=\itemindent,labelsep=0cm, align=left, itemsep=0.0cm, listparindent=0.5cm]
        \item A flag hyperedge $e_f$ may have the same syndrome bits as a normal hyperedge $e_n$, but may affect different Pauli frames. That is $\sigma(e_f) = \sigma(e_n)$, yet $\lambda(e_f) \neq \lambda(e_n)$.
        \item Two flag error events $e_f$ and $e_g$ may flip the same syndrome bits but may flip different flag bits. That is $\sigma(e_f) = \sigma(e_g)$ but $f(e_f) \neq f(e_g)$.
    \end{enumerate}
    To handle these situations, we propose categorizing hyperedges into \textit{equivalence classes}. Two edges $e_i$ and $e_j$ reside in the same equivalence class $C$ if $\sigma(e_i) = \sigma(e_j)$. Furthermore, during decoding, given a set of flag bits $F$, a single representative $\overline{e}$ is chosen from $C$ such that $|f(\overline{e}) \oplus F|$ is minimized. Furthermore, if $|F| > 0$, $\pi(\overline{e})$ is renormalized as in Equation~\ref{eqn:flagrenorm}, where $p_M$ is the measurement error probability. 
    Thus, we select the most probable error event from each equivalence class given a set of flag bits.
    \begin{equation}
        \label{eqn:flagrenorm}
        \pi(\overline{e}) \rightarrow p_M^{|f(\overline{e}) \oplus F|} \pi(\overline{e})^{|\sigma(\overline{e})| - 1}
    \end{equation}

    We briefly provide an example of how to leverage error equivalence classes. Consider the three error classes, $C_1$, $C_2$, and $C_3$, shown in Table~\ref{tab:equivclassex}. We run through three possible syndromes we might obtain during syndrome extraction.
    \begin{enumerate}[leftmargin=0.0cm,itemindent=0.5cm,labelwidth=\itemindent,labelsep=0cm, align=left, itemsep=0.15cm, listparindent=0.5cm, topsep=0.15cm]
        \item[\textbf{1: Syndrome $= \{ \sigma_0, \sigma_3 \}$, no flags. }] As there are no flags, we only use the error events from $C_1$ and $C_2$ that have no flags: $e_1=\{\sigma_0,\sigma_1,\sigma_2\}$ from $C_1$ and $e_2=\{\sigma_1,\sigma_2,\sigma_3\}$. A decoder should identify that $e_1$ and $e_2$ have occurred, as $\sigma(e_1) \oplus \sigma(e_2) = \{\sigma_0,\sigma_3\}$.
        \item[\textbf{2: Syndrome $=\{\sigma_0, \sigma_3\}$, Flags $=\{f_1\}$. }] Now that we have measured a flag bit, we must select events from each class whose flag bits are closest in similarity to $f_1$. From $C_1$, we select $e_1=\{\sigma_0,\sigma_1,\sigma_2\}, \{\}$ as it is only one flag off. From $C_2$, we select $e_2=\{\sigma_1,\sigma_2,\sigma_3\}, \{\}$ as it is the only option. From $C_3$, we select the flag edge $e_3=\{\sigma_0, \sigma_3\}, \{f_1\}$. A decoder should identify that $e_3$ has occurred as $\sigma(e_3) = \{\sigma_0,\sigma_3\}$/=.
        \item[\textbf{3: Syndrome $=\{\sigma_0,\sigma_1,\sigma_2\}$, Flags $=\{f_2,f_3\}$. }] From $C_1$, we select $\{\sigma_0,\sigma_1,\sigma_2\}, \{f_1,f_2,f_3\}$. From $C_2$, we select $\{\sigma_1,\sigma_2,\sigma_3\},\{\}$. From $C_3$, we select $\{\sigma_0\,\sigma_3\}, \{f_2,f_3\}$. A decoder should identify that $e_1$ has occurred.
    \end{enumerate}

    \begin{table}[!htb]
        \centering
        \begin{center}
        \caption{Example of Error Equivalence Classes}
        \label{tab:equivclassex}
        \begin{tabular}{|c|c|c|}
            \hline
            Class $C_1$ & Class $C_2$ & Class $C_3$ \\
            \hline
            \hline
            $\{ \sigma_0, \sigma_1, \sigma_2 \}, \{\}$ &
                \multirow{2}{*}{$\{ \sigma_1, \sigma_2, \sigma_3 \}, \{\}$} &
                $\{ \sigma_0, \sigma_3 \}, \{ f_1 \}$ \\
            $\{ \sigma_0, \sigma_1, \sigma_2 \}, \{ f_1,f_2,f_3 \}$ &&
                $\{ \sigma_0, \sigma_3 \}, \{ f_2, f_3 \}$ \\
            \hline
        \end{tabular}
        \end{center}
    \end{table}

\subsection{Decoding Hyperbolic Surface Codes}
    Our MWPM decoder for hyperbolic surface codes operates as follows. First, the decoding hypergraph must be translated into a decoding \textit{graph}. To do so, we collect representatives $\overline{e}$ from each error equivalence class. Then, for each $\sigma_i, \sigma_j \in \sigma(\overline{e})$, an edge $(\sigma_i, \sigma_j)$ is created in the decoding graph and is assigned a weight $w_{ij} = -\log \pi(\overline{e})$. Figure~\ref{fig:decoderex}(a) shows an example of this translation.

    \begin{figure}
        \centering
        \includegraphics[width=\columnwidth]{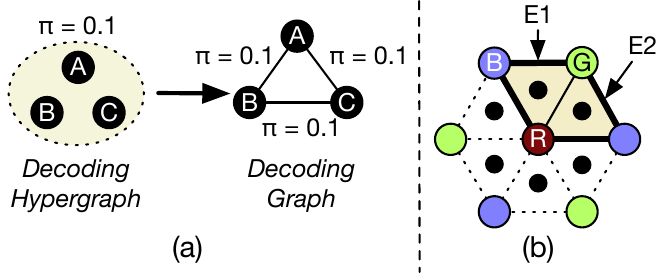}
        \caption{(a)~Example of a hyperedge between vertices $A$, $B$, and $C$ being translated to three edges. (b)~The edges from matching (bolded) are used to identify hyperedges $E1$ and $E2$ by lifting.}
        \label{fig:decoderex}
    \end{figure}

    Given the decoding graph, we must form a fully connected graph $G$, where vertices in $G$ are flipped syndrome bits. Each pair of flipped syndrome bits $(\sigma_i, \sigma_j)$ is assigned an edge whose weight $w_{ij}$ is equal to the weight of the shortest path between $\sigma_i$ and $\sigma_j$ in the decoding graph. Conceptually, this path corresponds to the most probable set of errors causing $\sigma_i$ and $\sigma_j$ to flip. Next, we pair all flipped syndrome bits to minimize the sum $\sum_{\sigma_i \leftrightarrow \sigma_j} w_{ij}$ to form a minimum-weight perfect matching. 
    
    To identify errors, if $\sigma_i$ and $\sigma_j$ are matched, we retrieve all edges in the path between $\sigma_i$ and $\sigma_j$ in the decoding graph. Usually, an edge $x$ in the path is present in the decoding hypergraph: in this situation, we update all Pauli frames $\lambda(x)$. However, when $x$ does not correspond to any edge, it is a flag edge, so we select the most similar flag edge $x_f$ and update all Pauli frames $\lambda(x_f)$.

\subsection{Decoding Hyperbolic Color Codes}
\label{sec:dechycc}
    Our Restriction decoder for hyperbolic color codes operates as follows. For the color codes, each syndrome bit $\sigma_i$ is associated with some color $C(\sigma_i) \in \{R, G, B\}$. Using these colors, we define three decoding graphs called \textit{restricted lattices}: $L_{RG}$, $L_{RB}$ and $L_{GB}$, which only contain syndrome bits of the specified colors. Then, we compute a minimum-weight perfect matching on these restricted lattices. 

    To identify errors, we must translate the matchings into errors on the color code 
    to identify errors. First, we collect all edges present in paths for each matching: we call this set $E_M$. If any flag edge $e_f$ appears twice in $E_M$, we immediately correct all Pauli frames $\lambda(e_f)$ and remove $e_f$ from $E_M$. This occurs if two syndrome bits are matched in different restricted lattices, and both matchings' paths contain $e_f$.
    
    All remaining edges are used in a lifting procedure; an example of lifting is shown in Figure~\ref{fig:decoderex}(b). First, all edges in $E_M$ are flattened such that their endpoints correspond to syndrome bits in the first round of syndrome extraction. This step handles measurement errors, which do not affect the logical state. Next, we identify all syndrome bits colored $R$ that are also incident on some edge in $E_M$. Then, for each incident syndrome bit, we select a maximal subset of incident edges from $E_M$: for each hyperedge $e$ outlined by this subset, we correct all Pauli frames $\lambda(e)$. We repeat the lifting procedure until $E_M$ is empty.

\ignore{
\subsection{Impact of Flag Qubits}
    Figure~\ref{fig:hyscflagber} compares the $X$ BER for a $[[60, 8, 6, 4]]$ hyperbolic surface code with and without flags using our MWPM decoder. Without flags, the BER is orders of magnitude worse. Furthermore, we find that the BER without flags scales $O(p^2)$, implying $d_\mathrm{eff} = 4$. In contrast, the BER with flags scales $O(p^3)$, indicating $d_\mathrm{eff} = 6$.

    \begin{figure}[!htb]
        \centering
        \includegraphics[width=0.9\columnwidth]{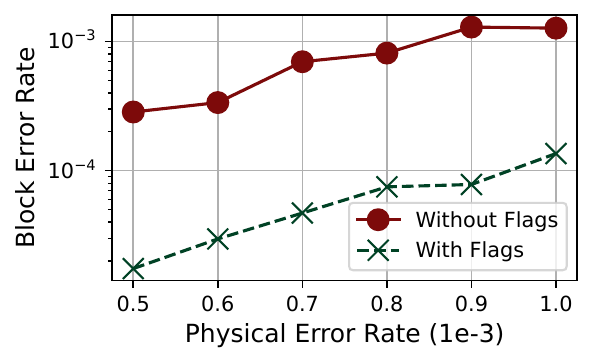}
        \caption{$X$ block error rates for a $[[60, 8, 6, 4]]$ hyperbolic surface code with and without flag qubits.}
        \label{fig:hyscflagber}
    \end{figure}
}

\subsection{Performance of Flagged Decoders}
\begin{table}[!htb]
    \vspace{-0.15in}
    \centering
    \begin{center}
    \caption{Evaluated Hyperbolic Codes}
    \label{tab:evaltbl}
    \begin{tabular}{|c|c||c|c|c|c|}
        \hline
        Family & Subfamily & $n$ & $k$ & $d_X$ & $d_Z$ \\
        \hline
        \hline
        \multirow{2}{*}{H. Surface Code} 
            & $\{4, 5\}$ & 160 & 18 & 8 & 6 \\
            & $\{5, 5\}$ & 150 & 32 & 6 & 6 \\
        \hline
        \multirow{2}{*}{H. Color Code}
            & $\{4, 6\}$ & 216 & 40 & 8 & 8 \\
            & $\{5, 8\}$ & 360 & 130 & 6 & 6 \\
        \hline
    \end{tabular}
    \end{center}
\end{table}
    Figure~\ref{fig:hyscber} and Figure~\ref{fig:hyccber} compare $BER_\mathrm{norm}$ for the hyperbolic surface and color codes in Table~\ref{tab:evaltbl} to the $d = 5$ and $d = 7$ planar surface and (6.6.6) color codes, respectively. Our setup for the planar color code uses the flag protocol proposed by Chamberland et al.~\cite{chamberland2020colorcodeswithflags}.

    \begin{figure}[!htb]
        \centering
        \includegraphics[width=\columnwidth]{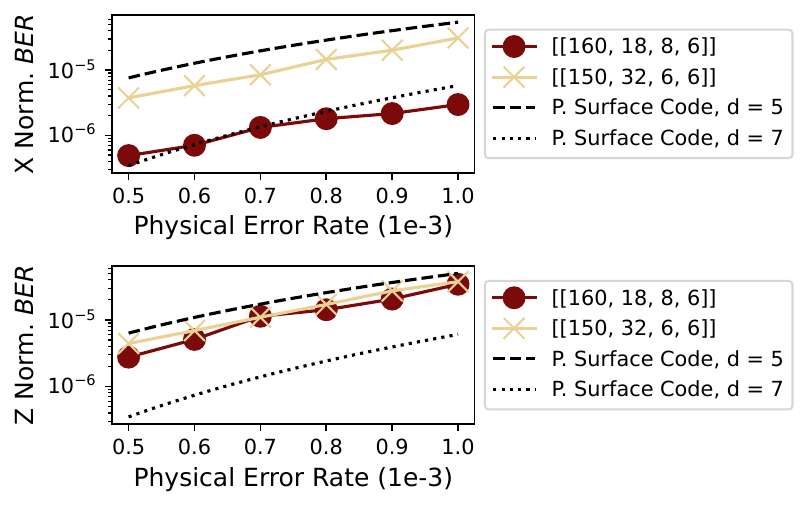}
        \vspace{-0.25in}
        \caption{$BER_\mathrm{norm}$ for surface codes.}
        \label{fig:hyscber}
        \vspace{-0.15in}
    \end{figure}

    \begin{figure}[!htb]
        \centering
        \includegraphics[width=\columnwidth]{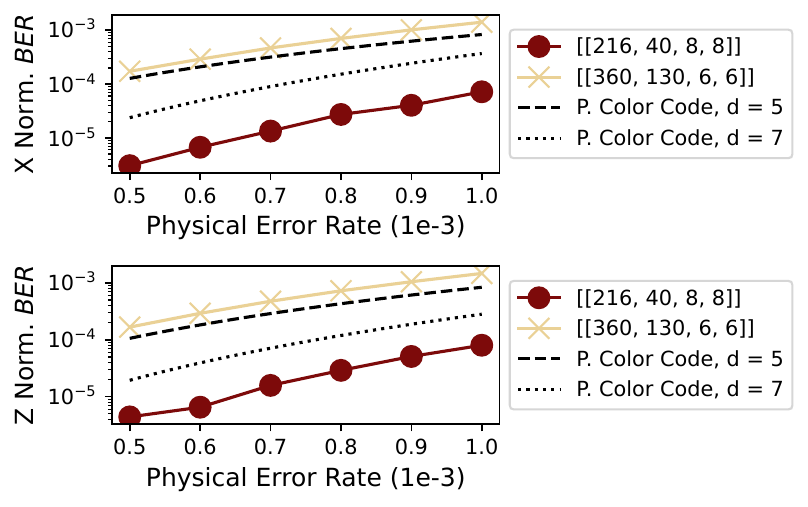}
        \vspace{-0.25in}
        \caption{$BER_\mathrm{norm}$ for color codes.}
        \label{fig:hyccber}
        \vspace{-0.15in}
    \end{figure}

    We find that the hyperbolic codes have competitive error rates with their planar counterparts while having higher $R_\mathrm{eff}$. In particular, the $[[150, 32, 6, 6]]$ hyperbolic surface code requires $424$ physical qubits to encode $32$ logical qubits while having comparable error rates to the $d = 5$ planar surface code, which would require $1568$ physical qubits. Similarly, the $[[216,40,8,8]]$ hyperbolic color code requires $512$ physical qubits to encode $40$ logical qubits while having comparable error rates to the $d = 7$ planar color code, which would require from $2200$ to $4000$ physical qubits, depending on implementation~\cite{landahl2011colorcodes, chamberland2020colorcodeswithflags}.

\subsection{Comparison with Prior Decoders}
    We briefly compare the proposed flagged decoders with prior work for planar codes. For hyperbolic surface codes, we compare against PyMatching~\cite{higgott2023sparseblossom}, an open source MWPM decoder for planar surface codes that has been used in much prior work~\cite{wu2022surfstitch, yin2023codestitch, higgott2023hyperbolicfloquetcodes, gidney2023yokedsc}. For hyperbolic color codes, we compare against Chromobius~\cite{gidney2023chromobius}, a recent open-source planar color code decoder, and Chamberland et al.'s Restriction decoder~\cite{chamberland2020colorcodeswithflags}, a planar color code decoder that uses flag qubits. As PyMatching and Chromobius do not work with flag qubits, we test them on architectures where parity qubits are directly connected to data qubits. Furthermore, as Chamberland et al.'s original work did not consider a generalized flag protocol, we modified the decoder to use our flag protocol during decoding.

    \begin{figure}[!htb]
        \centering
        \includegraphics[width=\columnwidth]{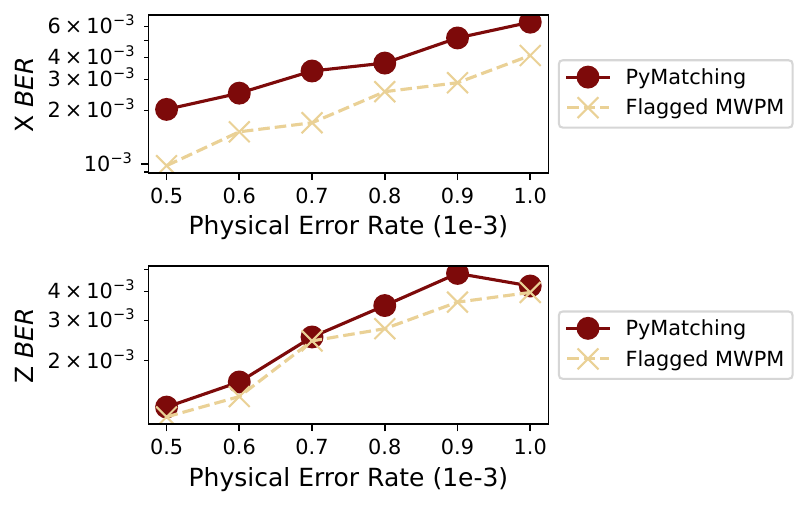}
        \vspace{-0.25in}
        \caption{$BER$ of a $[[30,8,3,3]]$ hyperbolic surface code using PyMatching and the flagged MWPM decoder.}
        \label{fig:mwpmdeccmp}
        \vspace{-0.15in}
    \end{figure}

    \subsubsection{Surface Codes} In this section, we consider a $[[30,8,3,3]]$ hyperbolic surface code from the $\{5,5\}$ subfamily. Figure~\ref{fig:mwpmdeccmp} shows the $X$ and $Z$ $BER$ for PyMatching and our flagged MWPM decoder. When handling $Z$ errors, PyMatching only achieves $d_\mathrm{eff} = 2$, whereas our flagged MWPM decoder achieves the full code distance ($d_\mathrm{eff} = 3$) by leveraging flag measurements.

    \subsubsection{Color Codes} In this section, we consider a $[[24,8,4,4]]$ hyperbolic color code from the $\{4,6\}$ subfamily. Chromobius, unfortunately, cannot decode the syndrome extraction circuit as it cannot handle error events where two parity qubits of the same color are flipped due to a CNOT propagation error. We also find that Chromobius cannot accurately decode when CNOT errors are disabled. Thus, Figure~\ref{fig:ccdeccmp} shows the $X$ and $Z$ $BER$ only for Chamberland et al.'s decoder and our flagged Restriction decoder. Chamberland et al.'s decoder only achieves $d_\mathrm{eff} = 2$, whereas the flagged Restriction decoder achieves the full code distance ($d_\mathrm{eff} = 4$).

    Given that the flag protocol used by both decoders is the same, we find that the flagged Restriction decoder outperforms Chamberland et al.'s decoder because Chamberland et al.'s decoder only handles flag edges in the MWPM stage of the decoder. In contrast, we handle flag edges outside the MWPM stage. Specifically, much of the improvement comes from where we handle flag edges $e_f$ that appear twice in $E_M$, as stated in Section~\ref{sec:dechycc}.
\section{Related Work}
\subsection{Architectural Construction}
    To the best of our knowledge, the only prior work in this area is by Tremblay et al. concerning architectures for \textit{Hypergraph Product (HGP)} codes~\cite{tremblay2022thinplanarconnectvity, tillich2013hgpcodes}. We note that the architecture considered in their paper is limited to HGP codes and has dense connectivity (at most degree-8 for the codes considered in their paper). Furthermore, using the authors' proposal, it is unclear how to support fault-tolerant syndrome extraction with flag qubits. 

\begin{figure}[!t]
    \centering
    \includegraphics[width=\columnwidth]{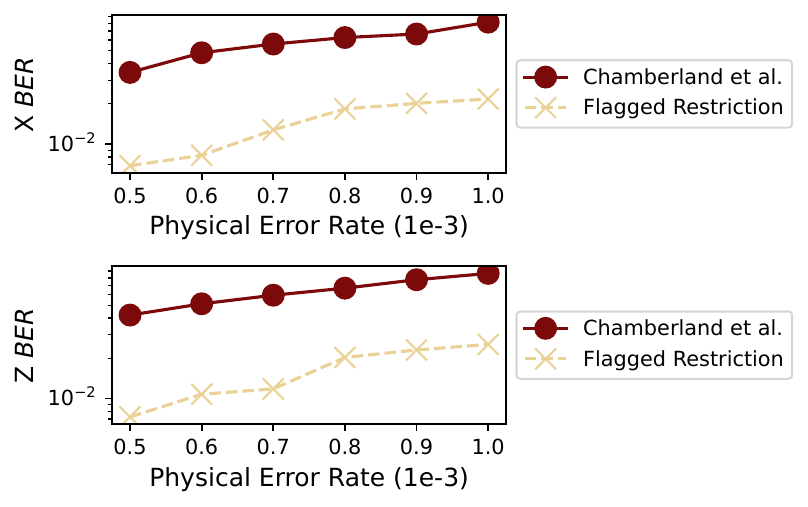}
    \vspace{-0.25in}
    \caption{$BER$ of a $[[24,8,4,4]]$ hyperbolic color code using Chamberland et al.'s decoder and the flagged Restriction Decoder.}
    \label{fig:ccdeccmp}
\end{figure}

\subsection{Code Mapping}
    Code mapping compilers, namely \textit{Surf-Stitch}~\cite{wu2022surfstitch} and \textit{Code-Stitch}~\cite{yin2023codestitch}, seek to map quantum codes to pre-existing quantum processors. The primary limitation of both these works concerns their support for fault-tolerant syndrome extraction. \textbf{Surf-Stitch} exclusively uses flag qubits to meet connectivity requirements, but this approach falls victim to the flag overuse problem. It also does not offer a flag protocol for handling flag syndromes, and thus, the results observed in the paper indicate that the mapped codes are not fault-tolerant. \textbf{Code-Stitch} has similar limitations regarding flag qubits but can also use Shor-style syndrome extraction to avoid using flag qubits. Shor-style syndrome extraction is fault-tolerant but has high overheads as measuring a weight-$\delta$ check requires $\delta$ parity qubits.

    In contrast, this paper proposes using proxy qubits to avoid the flag overuse problem and presents a flag protocol to recover the code distance. Nevertheless, note that the goal of code mapping compilers is moreso experimental, as they enable testing quantum codes on systems that do not support their connectivity requirements. In contrast, our goal focuses on relaxing the connectivity demands of quantum codes to facilitate the production of new processors.

\section{Conclusion}
    QLDPC codes are a scalable alternative to planar surface code. However, the biggest obstacles towards realizing these codes are practical, namely (1)~dense connectivity requirements greater than degree-4, (2)~fault-tolerant syndrome extraction circuits, and (3)~accurate decoding under circuit-level noise.
    For the first problem, we propose \textit{Flag-Proxy Networks (FPNs)} as a general architecture that uses flag and proxy qubits to achieve low connectivity while supporting fault-tolerant syndrome extraction. For the second, we propose a greedy scheduling algorithm generalizable to any quantum code. For the third, we propose a flag protocol to correct ``propagation errors" during syndrome extraction. Our evaluations on hyperbolic surface and color codes indicate that FPNs of these codes are respectively $2.9\times$ and $5.5\times$ more space-efficient than the $d = 5$ planar surface code. The error rates of these codes are also comparable to their planar counterparts.

\section*{Acknowledgement}
\noindent
This research was conducted using the {\em Partnership for an Advanced Computing Environment (PACE)} cluster at Georgia Tech. We thank Poulami Das (UT Austin) for her feedback on an earlier version of this manuscript.

\section*{Appendix I: Proof of Theorem 1}
\label{app:sec:proofoftheorem}
\noindent
\textbf{Theorem~\ref{th:faulttolerance}.} \textit{Suppose that a Flag-Proxy Network without proxies is fault-tolerant. The same network with proxies is also fault-tolerant.}
\begin{proof}
   There are two locations where proxy qubits may be added into an FPN: between the data and flag qubits and between the flag and proxy qubits. For brevity, we consider the first location, as our argument for the second is similar.

    \begin{figure}[!htb]
        \centering
        \vspace{-0.1in}
        \includegraphics[width=0.8\columnwidth]{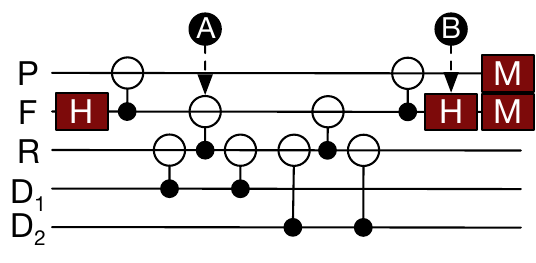}
        \vspace{-0.05in}
        \caption{The circuit used in the proof of Theorem~\ref{th:faulttolerance}.}
        \vspace{-0.05in}
        \label{fig:fpnproofcircuit}
    \end{figure}

    Consider the $Z$ syndrome extraction circuit shown in Figure~\ref{fig:fpnproofcircuit}, where data qubits $D_1$ and $D_2$ are in the joint state $\ket{D_1,D_2} = \alpha_{00}\ket{00} + \alpha_{01}\ket{01} + \alpha_{10}\ket{10} + \alpha_{11}\ket{11}$. $P$, $F$, and $R$ are parity, flag, and proxy qubits, respectively. At the start of the circuit, the state is as in Equation~\eqref{eqn:origstate}. The ideal evolution until \bcircled{A} yields the state in Equation~\eqref{eqn:afterpointa.ideal}. Now, without loss of generality, suppose that $CNOT(R, F)$ causes $Z$ errors on both $R$ and $F$. Then, we obtain the state in Equation~\eqref{eqn:afterpointa.faulty}.

    \noindent
    \begin{footnotesize}
    \begin{equation}
        \label{eqn:origstate}
    \begin{split}
        \ket{D_1, D_2}\ket{R}\ket{F, P} &=
            \left( \alpha_{00}\ket{00} + \alpha_{01}\ket{01} + \alpha_{10}\ket{10} + \alpha_{11}\ket{11} \right) \\
                &\quad\otimes
            \ket{0}
                \otimes
            \frac{\ket{00} + \ket{11}}{\sqrt{2}}
    \end{split}
    \end{equation}

    \begin{equation}
        \label{eqn:afterpointa.ideal}
        \ket{D_1, D_2, R, F, P} =
            \frac{ 
                \begin{split}
                \alpha_{00}\ket{00000} + \alpha_{00}\ket{00011} \\+ \cdots + \alpha_{11}\ket{11110} + \alpha_{11}\ket{11101} 
                \end{split}
            }{\sqrt{2}}
    \end{equation}
    \begin{equation}
        \label{eqn:afterpointa.faulty}
        \ket{D_1, D_2, R, F, P} =
            \frac{ 
                \begin{split}
                \alpha_{00}\ket{00000} - \alpha_{00}\ket{00011} \\+ \cdots + \alpha_{11}\ket{11110} - \alpha_{11}\ket{11101} 
                \end{split}
            }{\sqrt{2}}
    \end{equation}
    \end{footnotesize}
    
    The subsequent evolution until \bcircled{B} yields the state in Equation~\eqref{eqn:afterpointb}. Although a $Z$ propagation error has occurred on $D_1$ and $D_2$, it is detectable as measuring $F$ will always yield 1.
    
    \noindent
    \begin{footnotesize}
    \begin{equation}
        \label{eqn:afterpointb}
            \begin{split}
        \ket{D_1, D_2, F, P} &= 
                \alpha_{00}\ket{0010} - \alpha_{01}\ket{0111} \\
                &\quad- \alpha_{10}\ket{1011} + \alpha_{11}\ket{1111}
            \end{split}
    \end{equation}
    \end{footnotesize}

    Note that the propagation error resulting from the CNOT error on $R$ would have occurred even if it did not exist. Syndrome extraction with $R$ is functionally equivalent to syndrome extraction without $R$; including $R$ only increases the ``effective CNOT error" in the circuit. Hence, our argument above will also extend to fault-tolerant syndrome extraction circuits without flags, as the propagation error would not have harmed the effective distance. Therefore, adding proxies into an already fault-tolerant FPN does not reduce the effective distance.
\end{proof}

\section*{Appendix II: List of Codes}
    Tables~\ref{tab:hysctbl} and \ref{tab:hycctbl} list all hyperbolic codes evaluated in this paper. These codes were generated with code written using \textit{GAP}~\cite{GAP4}, a computer-algebra system, and the code distances were computed through brute-force search in \textit{Stim}~\cite{gidney2021stim}. Codes were generated according to the procedures described by Breuckmann et al. for hyperbolic surface codes~\cite{breuckmann2016constructionsofhyperbolicsurfacecodes} and by Higgott and Breuckmann~\cite{higgott2023hyperbolicfloquetcodes} for hyperbolic color codes.
    \begin{table}[!h]
        \centering
        \begin{center}
        \caption{List of Hyperbolic Surface Codes}
        \vspace{-0.1in}
        \label{tab:hysctbl}
        \footnotesize
        \begin{tabular}{|c||c|cccc|c|}
            \hline
            Subfamily & $R_\mathrm{ideal}$ & $n$ & $k$ & $d_X$ & $d_Z$ & $R_\mathrm{eff} (\%)$ \\
            \hline
            \hline
            \multirow{6}{*}{$\{4,5\}$} & \multirow{6}{*}{$1/10 = 0.1$}
                    & 60 & 8 & 6 & 4 & 4.3 \\
                  & & 160 & 18 & 8 & 6 & 3.6 \\
                  & & 360 & 38 & 8 & 8 & 3.3 \\
                  & & 660 & 68 & 10 & 8 & 3.3 \\
                  & & 1800 & 182 & 10 & 10 & 3.2 \\
                  & & 1920 & 194 & 12 & 10 & 3.2 \\
            \hline
            \multirow{3}{*}{$\{4,6\}$} & \multirow{3}{*}{$1/6 \approx 0.17$} 
                    & 36 & 8 & 4 & 4 & 7.2 \\
                  & & 336 & 58 & 8 & 6 & 5.5 \\
                  & & 864 & 146 & 10 & 8 & 5.3 \\
            \hline
            \multirow{5}{*}{$\{5, 5\}$} & \multirow{5}{*}{$1/5 = 0.2$}
                    & 30 & 8 & 3 & 3 & 9.4 \\
                  & & 40 & 10 & 4 & 4 & 9.3 \\
                  & & 80 & 18 & 5 & 5 & 8.0 \\
                  & & 150 & 32 & 6 & 6 & 7.5 \\
                  & & 900 & 182 & 8 & 8 & 7.2 \\
            \hline
            \multirow{3}{*}{$\{5,6\}$} & \multirow{3}{*}{$4/15 \approx 0.27$}
                    & 60 & 18 & 4 & 3 & 10.6\\
                &   & 120 & 34 & 6 & 5 & 10.0 \\
                &   & 2520 & 674 & 8 & 6 & 9.2 \\
            \hline
        \end{tabular}
        \end{center}
    \end{table}
    
    \begin{table}[!h]
        \centering
        \vspace{-0.25in}
        \begin{center}
        \caption{List of Hyperbolic Color Codes}
        \label{tab:hycctbl}
        \footnotesize
        \begin{tabular}{|c||c|ccc|c|}
            \hline
            Subfamily & Asymptotic Rate & $n$ & $k$ & $d$ & $R_\mathrm{eff} (\%)$ \\
            \hline
            \hline
            \multirow{5}{*}{$\{4,6\}$} & \multirow{5}{*}{$1/6 \approx 0.17$} 
                    & 24 & 8 & 4 & 15.1 \\
                  & & 120 & 24 & 6 & 8.3 \\
                  & & 216 & 40 & 8 & 7.8\\
                  & & 1320 & 224 & 10 & 7.0 \\
                  & & 1440 & 244 & 12 & 6.8\\
            \hline
            \multirow{3}{*}{$\{4, 8\}$} & \multirow{3}{*}{$1/4 = 0.25$}
                    & 32 & 12 & 4 & 17.1 \\
                  & & 400 & 104 & 8 & 11.1 \\
                  & & 2688 & 676 & 12 & 10.6 \\ 
            \hline
            \multirow{2}{*}{$\{4, 10\}$} & \multirow{2}{*}{$3/10 = 0.3$}
                    & 40 & 16 & 4 & 18.2 \\
                  & & 1000 & 304 & 8 & 12.5 \\
            \hline
            \multirow{3}{*}{$\{5, 8\}$} & \multirow{3}{*}{$7/20 = 0.35$}
                    & 320 & 116 & 4 & 14.6 \\
                  & & 360 & 130 & 6 & 14.3 \\
                  & & 1920 & 676 & 8 & 14.0 \\
            \hline
        \end{tabular}
        \end{center}
    \end{table}
    
    \noindent
    \textbf{Note:} We found the Restriction decoder (and variants like the M\"{o}bius decoder~\cite{sahay2022mobius, gidney2023chromobius}) cannot accurately decode several hyperbolic color codes under code capacity noise (no operation error). We often found that such codes often have a lower distance counterpart with the same $n$ and $k$. However, a concurrent work, the \textit{Concatenated MWPM} decoder~\cite{lee2024concatmwpm}, can accurately decode most hyperbolic color codes (under code capacity noise). However, the decoder needs our flag protocol to achieve the full distance of the hyperbolic color codes under circuit-level noise. As a result of this phenomenon, Table~\ref{tab:hycctbl} only contains hyperbolic color codes that the Restriction decoder can accurately decode under code capacity noise. 
    
    \vspace{0.05in}
    \noindent
    \textbf{Planarity:} All FPNs listed above are \textit{biplanar}, much like bivariate bicycle codes~\cite{bravyi2023grosscodes} and the hyperbolic floquet codes~\cite{higgott2023hyperbolicfloquetcodes}.
    
%
%
%
%
%

\section*{Appendix III: Artifact}

\subsection{Abstract}

Our artifacts are the codebase which is used to construct and evaluate FPNs. We have also provided the codes that are used in our evaluations.

Our artifact allows the user to reproduce the following data: Figure~10(a), Figure~12, Table~I, Figure~14, Figure~17, Figure~18, Figure~19, and Figure~20, which are all the quantitative results of our paper. After running the requisite experiments, all data is viewable in a provided iPython notebook.

\subsection{Artifact check-list (meta-information)}

{\small
\begin{itemize}
  \item {\bf Program:} {\footnotesize\texttt{protean}, \texttt{pr\_base\_memory}, \texttt{pr\_planar\_memory}}
  \item {\bf Compilation:} \texttt{gcc}, at least version 12
  \item {\bf Data set:} Hyperbolic quantum codes provided by the authors.
  \item {\bf Hardware:} FPN generation requires MacOS. BER evaluations require a computing cluster.
  \item {\bf Execution:} Python and bash scripts which automate the experiments.
  \item {\bf How much disk space required (approximately)?:} At most 4GB
  \item {\bf How much time is needed to prepare workflow (approximately)?:} At most 30 minutes
  \item {\bf How much time is needed to complete experiments (approximately)?:} At most 3 days
  \item {\bf Publicly available?:} Yes
  \item {\bf Code licenses (if publicly available)?:} MIT
  \item {\bf Data licenses (if publicly available)?:} MIT
  \item {\bf Workflow automation framework used?:} CMake, version 3.20.2 or higher
  \item {\bf Archived (provide DOI)?: }10.5281/zenodo.13325358
\end{itemize}
}

\subsection{Description}

\subsubsection{How to access}

Available on Zenodo \href{https://zenodo.org/doi/10.5281/zenodo.13325358}{here}.

\subsubsection{Hardware dependencies}
    To replicate the construction of the \textit{Flag-Proxy Networks (FPNs)}, a laptop with MacOS is needed: such evaluations are expected to take no more than six hours. To replicate \textit{Block Error Rate (BER)} evaluations, a computing cluster will be necessary to handle large codes. Small codes can be evaluated on any laptop.

    Note that while FPN generation can be "run" on Linux, we found that there is a bug on Linux that prevents syndrome extraction circuits from containing CNOTs: it is unclear why this bug occurs (culprit is likely in the scheduling code).

\subsubsection{Software dependencies}
    Our code compiles with \texttt{gcc-12} through \texttt{gcc-14}. It has not been tested with \texttt{clang}. Furthermore, our codebase uses the CMake build tool (version 3.20.2 or higher) to automate compilation.
    
    Constructing FPNs requires installing CPLEX, which is free for academics (see \href{https://www.ibm.com/academic/}{here}). If it is not possible to install CPLEX, we can provide the FPNs for all evaluated codes. On the other hand, evaluating BERs requires MPI to parallelize these evaluations across many cores on large clusters. The MPI version must match the corresponding \texttt{gcc} version. When building our code with \texttt{gcc-14}, we were able to build our code using \texttt{openmpi} v5.0.3. All other dependencies, such as Stim, are packaged with our codebase.

    Finally, to execute the experiments, we have provided Python scripts which execute the scripts on the requisite quantum codes and also set the appropriate flags for the program. To run these scripts, we require Python 3.10 or higher. To create the plots used in our paper, we require \texttt{matplotlib} v3.8.3, \texttt{numpy} v1.26.3, \texttt{scipy} v1.11.4, and a method of opening an iPython notebook (i.e. JupyterLab).

\subsubsection{Data sets}
    We have provided representations of the evaluated quantum codes in the folder \texttt{data/tanner}. Hyperbolic surface codes are listed under \texttt{hysc} and hyperbolic color codes are listed under \texttt{hycc}. Furthermore, both code folders are organized by sub-family.

    These codes are constructed using \textit{GAP}, a free and commonly-used computer-algebra system. We do not include the corresponding GAP scripts in the artifact to reproduce constructing the codes, as the code construction is rather tedious. If you require these scripts, please reach out to the corresponding author.

\subsection{Installation}

\subsubsection{CPLEX Configuration}
    Making FPNs relies on linking IBM's CPLEX solver to some of the built executables. Currently, we have an automated method of finding CPLEX on MacOS devices in the CMake file \texttt{cmake/FindCPLEX.cmake}. This method may not work on a Windows or Linux machine, but the corresponding CMake variables, \texttt{CPLEX\_LIBRARY\_DIR} and \texttt{CPLEX\_INCLUDE\_DIR}, may be
    set manually by modifying their definition in this file, or providing the paths when running CMake (see below). However, we do note that our FPN generation program only works on MacOS. Please reach out to the authors for the FPNs produced in the paper.

    If you are using different machines to make FPNs and run memory experiments (as we did since our cluster did not support CPLEX), then on the machine that does not support CPLEX, set the variables \texttt{COMPILE\_PROTEAN\_LIB} and \texttt{COMPILE\_PROTEAN\_MAIN} to \texttt{OFF} in the file \texttt{cmake/UserConfig.cmake}.

\subsubsection{Building Executables}
    Run the following commands to build the necessary executables:

    \vspace{0.1in}
    \noindent
    {\footnotesize 
    \texttt{\$ mkdir Release \&\& cd Release \\
    \$ cmake .. -DCMAKE\_BUILD\_TYPE=Release [-DCPLEX\_LIBRARY\_DIR=... -DCPLEX\_INCLUDE\_DIR=...] \\
    \$ make -j4}}

\subsection{Experiment workflow}
    \subsubsection{Figure~10, Table~I, Figure~12, Figure~14}
        Building the FPNs can be done by running the bash script \texttt{\footnotesize scripts/protean/make\_all\_arch.sh}. This must be done in the base directory (same level as \texttt{Release}). This script will call the Python file \texttt{\footnotesize scripts/protean/evals.py}, which will build FPNs for all families and sub-families. 
        
        These experiments should take at most six hours on a laptop.

    \subsubsection{Figure~17, Figure~18}
    To run the main memory experiments (for the codes in Table~III), run the \texttt{\footnotesize scripts/protean/compute\_ber\_5e-4\_1e-3.sh} script as follows:
    
    \vspace{0.1in}
    \noindent
    {\footnotesize
    \texttt{\$ ./scripts/protean/compute\_ber\_5e-4\_1e-3.sh hysc/5\_5/150\_32\_6\_6 mwpm -mx \\
    \$ ./scripts/protean/compute\_ber\_5e-4\_1e-3.sh hysc/5\_5/150\_32\_6\_6 mwpm -mz }}
    \\
    
    Repeat the same commands for \texttt{hysc/4\_5/160\_18\_8\_6} as well. These commands will execute memory experiments for the hyperbolic surface codes in Table~III. For the hyperbolic color codes in Table~III, run the above commands for \texttt{hycc/4\_6/216\_40\_8\_8} and \texttt{hycc/5\_8/360\_130\_6\_6} but also replace \texttt{mwpm} with \texttt{restriction}.

    We have provided our version of the \texttt{\footnotesize scripts/protean/compute\_ber\_5e-4\_1e-3.sh}, which is designed to work with our compute cluster. We have commented out lines that correspond to module imports (i.e. for OpenMPI). Note that our version uses \texttt{srun}, which implicitly calls \texttt{mpirun}, as our cluster uses SLURM. Hence, if the user's cluster does not use SLURM, it will need to be switched out (i.e. to \texttt{mpirun <X>}, where $X$ is the number of cores used).

    For each code, we used 512 cores, 8GB of memory per core, and a 12 hour wall-time.

\subsubsection{Figure~19, Figure~20}
    Run the following bash script: \texttt{\footnotesize ./scripts/protean/eval\_decoders.sh <X>}, where $X$ is the number of cores used. This can be done on a laptop within a few minutes.

\subsection{Evaluation and expected results}
All results can be found by examining their respective cells in the iPython notebook \texttt{\footnotesize scripts/protean/plots.ipynb}. Furthermore, all figures can be found in \texttt{\footnotesize scripts/protean/plots} after running the respective cells in the notebook.

Generated plots and data should roughly match what is reported in the main text, with some possible variance due to randomness.



\subsection{Methodology}

Submission, reviewing and badging methodology:

\begin{itemize}
  \item \url{https://www.acm.org/publications/policies/artifact-review-and-badging-current}
  \item \url{https://cTuning.org/ae}
\end{itemize}

\bibliographystyle{IEEEtranS}
\bibliography{refs}

\end{document}